\documentstyle[preprint,amsfonts,amssymb,aps,12pt]{revtex}
\newcommand{\beq}{\begin{equation}}
\newcommand{\eeq}{\end{equation}}
\input{epsf}
\tighten
\begin{document}
\draft
%
%
\title{Monte Carlo Simulation of Long Chain Polymer Melts:\\
Crossover from Rouse to Reptation Dynamics}
\author{T. Kreer$^1$\footnote{To whom correspondence should be addressed.  
Email: {\sf tkreer@plato.physik.uni-mainz.de}}, J. Baschnagel$^2$, M. M\"uller$^1$, K. Binder$^1$\\[2mm]}
\address{$^1$Institut f\"ur Physik, Johannes-Gutenberg Universit\"at, D-55099 Mainz, Germany\\[1mm]}
\address{$^2$Institut Charles Sadron, 6 rue Boussingault, F-67083, Strasbourg Cedex, France}
\maketitle
%
%
%
%
\newcommand{\mr}[1]{{\rm #1}}
\renewcommand{\vec}[1]{\mbox{\boldmath$#1$\unboldmath}}
%
%
%
%
%
\begin{abstract}
We present data of Monte Carlo simulations for monodisperse linear polymer chains in dense melts with
degrees of polymerization between $N=16$ and $N=512$. The aim of this study is to investigate the 
crossover from Rouse-like dynamics for short chains to reptation-like dynamics for long chains. To
address this problem we calculate a variety of different quantities: standard mean-square displacements
of inner monomers and of the chain's center of mass, the recently proposed cubic invariant [U. Ebert
{\em et al.}, Phys.\ Rev.\ Lett.\ {\bf 78}, 1592 (1997).], persistence of bond-vector orientation with
time, and the auto-correlation functions of the bond vector, the end-to-end vector and the Rouse modes.
This analysis reveals that the crossover from non- to entangled dynamics is very protracted. Only
the largest chain length $N=512$, which is about 13 times larger than the entanglement length,
shows evidence for reptation.
\end{abstract}
%
%
\pacs{{\sf PACS}: 61.20.Ja,64.70.Pf,61.25.Hq,83.10.Nn\\%
submitted to {\em Macromolecules} on August {\bf 24}, 2000
}
%
%
\section{Introduction}
\label{intro}

A detailed description of the slow dynamics of entangled polymer chains would be of great interest for better understanding 
the physical properties of dense polymer melts or solutions.  Several attempts to solve this problem have been made. 
In a seminal paper de Gennes introduced the concept of reptation \cite{b2}. The fundamental idea is that the motion 
of a polymer is not spatially isotropic, but has to occur along the contour of a ``tube'' which is formed by the 
surrounding chains. These surrounding chains are treated as a fixed network of impenetrable obstacles. The simultaneous
motion of the polymers and possible correlations resulting from that are disregarded. The elaboration of this idea has 
led to a rich scenario of testable predictions \cite{b3,b,a}. Critical analysis of these predictions by experiments
\cite{j,c,r,c2,c3} and computer simulations \cite{d3,d,o} suggest that reptation is an important relaxation mechanism. 
However, it cannot provide a quantitative explanation of all dynamic features observed. These deviations could result from
additional relaxation processes which blur the clear signature of reptation for typical experimental chain lengths $N$ and
become negligible only in the asymptotic limit of extremely large $N$ (see, however, \cite{sasharubi} for a different 
asymptotic behavior).

{ 
These additional relaxation mechanisms could involve fluctuations of the tube's contour 
\cite{e,milnermcleish,k2}, reorganization of the tube \cite{b4} by release and creation of topological constraints 
\cite{j,e,b5}, elastic distortions of the entanglement network \cite{ro} or long-lived density fluctuations \cite{sasha}. 
All these extensions
assume the existence of a tube. None of them explains its microscopic origin. An attempt to develop a microscopic 
theory was made by the development of a mode-coupling theory for polymers \cite{ken1,ken2,schsz,i,fs_1997,fs_1997_jcp}. 
This approach leads to similar predictions as reptation theory without invoking a tube hypothesis.

However, some doubt was cast on this concept by simulations of a model by Shaffer\cite{BREF1} which respects excluded 
volume interactions between the monomers of the chains, but allows Monte Carlo moves that lead to intersections between 
bonds connecting the monomers. For long chains in this model a simple Rouse--like behavior is found. Only when the 
conditions that bonds must not intersect is used as an additional constraint in the dynamics, a reptation--like behavior 
results from the simulation. These results, where static correlations are precisely identical for both models, but one 
dynamic version exhibits Rouse behavior and the other reptation--like behavior, seem to imply that the topological 
constraint of the non--crossability of chains is not an automatic consequence of the excluded volume interaction. In 
this context, we note that many other attempts have been made to account for the slow dynamics of entangled polymer 
chains by concepts differing from reptation, e.g., collective motion of many entangled chains \cite{BREF2}, 
but we are not attempting to critically evaluate all these theories here.
}

In view of these diverse attempts to improve on reptation theory, it is also beneficial to consider a model which 
exhibits pure reptation behavior in the asymptotic limit of large chains and long times, and for which finite-$N$ and 
finite-time corrections can be calculated. Evans and Edwards proposed such a model in which a single chain moves 
through a regular array of impenetrable obstacles \cite{ee}. The obstacles do not influence the equilibrium 
configuration of the chain, but only the accessible relaxation moves of its monomers. This model has recently been 
reanalyzed analytically \cite{t,u1}. Detailed comparisons with computer simulations \cite{t,u,sbe} reveal an
extremely slow crossover to the asymptotic power laws characteristic of reptation. Even in the ideal situation,
where the tube diameter is equal to the bond length, the power laws only become clearly pronounced for $N > 100$. If
the lattice of obstacles is diluted, the tube diameter increases and the crossover to reptation is shifted to
larger chain lengths \cite{sbe}. 

The latter study resembles more closely the situation encountered in computer simulations of melts where the tube diameter is 
in general larger than the bond length. Taking into account that simulations typically work with $N < 1000$, the analysis 
of the Evans-Edwards model suggests that the results of polymer melt simulations are characteristic of the crossover from non-entangled, 
Rouse-like dynamics for short chains to (slightly) entangled, reptation-like dynamics for long chains. With the present 
study we want to further investigate this crossover by a Monte Carlo simulation of the bond-fluctuation model. Our paper is 
organized as follows: Section~\ref{modsta} introduces the model and describes some of its static properties. 
Section~\ref{dyn} represents the main part of the paper. It deals with dynamics of the model. Here, we compare various
mean-square displacements and orientational correlation functions, extract the chain length dependence of relaxation
times and of the diffusion coefficient, and estimate the parameters of reptation theory, such as the tube diameter or the
entanglement length. The final section~\ref{conc} summarizes our conclusions.

\section{Model and Static Properties}
\label{modsta}
The aim of the present work is to extend and to complement previous studies of the crossover
from Rouse to reptation dynamics using the bond-fluctuation model \cite{m,wittmer,mwb}. The
bond-fluctuation model is a coarse-grained lattice model \cite{o,kk,kb_kbrev}, in which a polymer
is represented by a chain consisting of $N$ monomers. Each monomer occupies a unit cell of a simple 
cubic lattice (unit cell = 8 lattice sites). This increased monomer size allows the bond 
vectors $\bbox{l}$ to vary in both length and direction to a much larger extent than in simpler 
lattice models where a monomer is associated with a single site \cite{kk}. In total, there are 108
bond vectors \cite{w} giving rise to 87 bond angles and 5 different bond lengths ($2,\sqrt{5},
\sqrt{6},3,\sqrt{10}$). The average bond length and bond angle depend on the volume fraction, 
$\phi = 8NP/L^3$, of occupied lattice sites ($P=$ total number of polymers in the simulation box, 
$L=$ linear dimension of the box). Choosing $\phi=0.5$ the model realizes typical properties 
of dense melts \cite{m}. For this density the mean bond length and mean cosine of the bond angle 
are given by $l=\langle \bbox{l}^2 \rangle^{1/2} = 2.64$ and $\alpha=\langle \cos \theta \rangle 
=-0.1055$, respectively \cite{bptb}.

To analyze the change in dynamic behavior with increasing $N$ simulations were performed for
different chain lengths ranging from $N=16$ to $N=512$ (see Table~\ref{tab1}). This extends 
previous studies \cite{m} by a factor of 2 only, but it amounts to an increase in relaxation time
by almost an order of magnitude (cf.\ Table~\ref{tab1}). Furthermore, the system size could also be 
chosen much larger. For most chain lengths the linear dimension of the simulation box is more
than a factor of 2 greater than the average end-to-end distance $R_{\text{e}}$, i.e., $L > 2 
R_{\text{e}}$ (with the exception of $N=512$ where $L=128 \simeq 1.75 R_{\text{e}}$ only;
see Table~\ref{tab1}). Since finite-size effects may be expected if $L \leq 2 R_{\text{e}}$,
Fig.~\ref{rerg} compares the mean-square end-to-end distance $R^2_{\text{e}}$ and mean-square 
radius of gyration $R^2_{\text{g}}$ of the present study with the values of earlier works 
\cite{m,wittmer}. Two conclusions can be drawn from this figure. First, finite-size effects
did not affect previous (static) results. Within the statistical uncertainties the old data for
$N=200$ ($R_{\text{e}} \simeq 45$, $L=40$ \cite{m}) agree with the present analysis. Second, the
bond-fluctuation model approaches the asymptotic ideal chain limit, i.e., $R^2_{\text{e}} = 6
R^2_{\text{g}} = C_\infty l^2 (N-1)$, if $N > 100$. Here, $C_\infty$
($= \lim_{N\rightarrow \infty} R^2_{\text{e}}/[(N-1)l^2]$) is the characteristic ratio which
measures the swelling of a chain due to its intrinsic stiffness compared to a pure random walk.
Using $C_\infty \simeq 1.56$ (see Fig.~\ref{rerg}) and $l_{\text{p}}=l(C_\infty+1)/2$ \cite{flory},
the persistence length, $l_{\text{p}}$, of the model is given by $l_{\text{p}} \simeq 3.38$.

The shape of a polymer is not spherical. A good indicator of its geometrical form is the 
gyration tensor {\sf Q} whose elements are given by \cite{s,s2,s3}
\beq 
Q_{\alpha \beta}=\frac{1}{N}\sum_{n=1}^N (R_{\alpha,n}-R_{{\rm cm},\alpha})
(R_{\beta,n}-R_{{\rm cm},\beta}) \qquad (\alpha,\beta=1,\ldots,3)\;.
\label{eq:gtensor}
\eeq
Here, $R_{\alpha,n}$ and $R_{{\rm cm},\alpha}$ denote the $\alpha$th-component of the position 
vectors to monomer $n$ and to the chain's center of mass, i.e., $\bbox{R}_n=(R_{1,n},R_{2,n},
R_{3,n})$ and $\bbox{R}_{\rm cm}=(R_{{\rm cm},1},R_{{\rm cm},2}, R_{{\rm cm},3})$. The trace
of this tensor provides another way to calculate the radius of gyration \cite{s}
\beq
\langle {\rm Tr}\, \mbox{{\sf Q}} \rangle = \langle \lambda_1 \rangle + \langle \lambda_2 \rangle 
+ \langle \lambda_3 \rangle= R^2_{\rm g} \;,
\label{lambda.eq}
\eeq
where $\lambda_\alpha$ are the eigenvalues of {\sf Q}. If the shape of a polymer was
spherical, one would have $\langle \lambda_1 \rangle=\langle \lambda_2 \rangle=\langle \lambda_3 \rangle$. 
Contrary to that, Fig.~\ref{lambda}
shows that the biggest eigenvalue, $\langle \lambda_1\rangle$, is much larger than the other two eigenvalues. 
This implies that the shape of a chain is distorted with respect to a sphere: It is stretched in 
direction of the 
long principal axis and shrunk in directions of the axes corresponding to $\langle \lambda_2\rangle$ and 
$\langle \lambda_3\rangle$. Therefore, a chain rather resembles a ``flattened American football''%
\footnote{Recent studies \cite{jtc,em} of the shape of Gaussian chains indicate that the visualization of
a random-walk polymer as a flattened American football is not completely correct. The density distribution
of monomers in the coordinate system of the principal axes exhibits a slight minimum at the origin (i.e.,
at the center of mass) for the largest axis, whereas it has a maximum for the other two axes. Therefore, 
the shape is rather dumbbell-like.}. This characteristic property can be exploited to construct an efficient 
coarse-grained model for dense polymer melts \cite{mk,cg_review}.

In the course of the simulation, a monomer and a lattice direction (out of 6) are selected at
random, and a jump by one lattice constant is attempted in that direction. The jump is
accepted if the resulting bond vectors belong to the allowed set of bonds and if the targeted
lattice sites are empty. Otherwise, the move is rejected. This random displacement simulates
the impact of a stochastic force which is exerted on a monomer by its local environment. 

In the single chain limit, this local jump model is expected to give rise to Rouse dynamics if 
hydrodynamic interactions are absent \cite{a}. Since hydrodynamic forces are mediated by solvent
molecules, an isolated chain of the bond-fluctuation model provides an example for this conjecture
\cite{o,mwb}. However, if the volume fraction $\phi$ increases, deviations from Rouse dynamics
are observed due to the mutual interaction of the chains and due to the onset of entanglements
\cite{o}. Nevertheless, the Rouse model is generally believed to provide a reliable description 
of short (non-entangled) chains in a melt. Therefore, we determined its basic variables, the Rouse
modes \cite{verdier}    
\beq
\bbox{X}_p(t)= \frac{1}{N}\sum_{n=1}^N \bbox{R}_n(t) \cos\frac {(n-1/2)p\pi}{N}
\quad (p=0,1,\ldots,N-1) \; ,
\label{rousemodedis.eq}
\eeq
where $\bbox{R}_n(t)$ is the position of the $n$th monomer at time $t$. At $t=0$ the (static)
correlation of the modes reflects the structural properties of the polymers. For ideal random
walk chains it is given by \cite{verdier}
\beq
\langle \bbox{X}_{p}(0)\cdot \bbox{X}_{q}(0)\rangle = \delta_{pq}\, \frac{b^2}{8N}\left[\frac{1}{
\sin (p\pi/2N)}\right]^{2} \; \stackrel{p/N \ll 1}{\longrightarrow} \;
\delta_{pq}\, \frac{1}{2\pi^2} \frac{Nb^2}{p^2} \qquad (\mbox{for $p>0$}) \;.
\label{phirou.eq}
\eeq
In this equation $b$ denotes the effective bond length defined by $b^2=C_\infty l^2$ \cite{a}.
However, we use $b^2 = R^2_{\text{e}}/(N-1)$. This ratio is (slightly) chain-length dependent.
It varies from about 8.7 for $N=16$ to about 10.6 for $N=512$, i.e., by about $18\%$, in our
simulation. 

Using this identification Eq.~(\ref{phirou.eq}) suggests that a plot of 
$8N(N-1)\langle \bbox{X}_{p}(0)^2\rangle/R^2_{\text{e}}$ versus $p/N$ should yield a master
curve. Figure~\ref{autokorr} shows a test of this prediction. In fact, the data for all chain 
lengths collapse onto a common curve which is nicely described by $[\sin(p\pi/2N) ]^{-2}$ if $p/N 
\lesssim 0.05$. However, if $p/N$ increases, deviations between the simulation
results and the Rouse prediction gradually develop. The Rouse theory overestimates the 
correlation, especially for $p/N \gtrsim 0.3$. These large modes probe local distances along the
backbone of a chain. For instance, $p/N \gtrsim 0.3$ corresponds to subunits of a trimer and
smaller than that. On these local scales, the intrinsic stiffness of a polymer should be taken into account. 

The simplest way to achieve this consists in replacing the random walk by a freely rotating chain
\cite{a,flory}. This is a chain model, in which the mean bond length and bond angle are fixed 
and each bond is allowed to rotate freely around the direction of the preceding bond. Starting 
from 
\[
\langle \bbox{X}_p^2(0) \rangle = \frac{1}{N^2} \sum_{n,m=1}^N \langle \bbox{R}_n(0) \cdot 
\bbox{R}_m(0) \rangle \cos \frac{p\pi (n-1/2)}{N} \cos\frac{p\pi(m-1/2)}{N} \;,
\]
this implies to write (remember $\alpha = \langle \cos \theta \rangle = - 0.1055$)
\[
\langle \bbox{R}_n \cdot \bbox{R}_m \rangle = \sum_{i=1}^{n-1} \sum_{j=1}^{m-1}
\langle \bbox{l}_i \cdot \bbox{l}_j \rangle = l^2\sum_{i=1}^{n-1} \sum_{j=1}^{m-1}
(-\alpha)^{|i-j|}
\]
instead of $\langle \bbox{l}_i \cdot \bbox{l}_j \rangle =l^2 \delta_{ij}$ for a random walk.
Using furthermore
\begin{eqnarray*}
\lefteqn{\sum_{n,m=2}^N \cos \frac{p\pi (n-1/2)}{N} \cos\frac{p\pi(m-1/2)}{N} \sum_{i=1}^{n-1} 
\sum_{j=1}^{m-1} (-\alpha)^{|i-j|}} \nonumber \\
&=&
\sum_{n=2}^N \sum_{i=2}^n \sum_{m=i}^N \cos \frac{p\pi (n-1/2)}{N} \cos\frac{p\pi(m-1/2)}{N} 
\left[ \frac{(-\alpha)^{m-i+1}-1}{(-\alpha)-1}+ \frac{(-\alpha)^{n-i+1}-1}{(-\alpha)-1} -1
\right] \;,
\end{eqnarray*}
one obtains
\begin{eqnarray}
\frac{8N(N-1)}{R^2_{\text{e}}} \langle \bbox{X}_{p}^2(0)\rangle 
&\simeq&
\left[\frac{1}{\sin (p\pi/2N)}\right]^{2} -
\frac{4(-\alpha)}{1-2(-\alpha)\cos(p\pi/N)+(-\alpha)^2} \; \times \nonumber \\
&&
\left [1 +\frac{1}{N}\frac{2\alpha (1+\alpha)}{1-\alpha} \frac{1}{1+2\alpha\cos(p\pi/N)+
\alpha^2} \frac{\sin (p\pi/N)} {\tan (p\pi/2N)} \right]\; ,
\label{rousetrimer}
\end{eqnarray}
where the $\simeq$--sign indicates that $|\alpha| = 0.1055 \ll 1$ was used to write the 
$1/N$-correction in the second line of Eq.~(\ref{rousetrimer}) in the present form. 
Although this correction violates the $p/N$-scaling, its numerical value is always smaller
than 0.03 (= ``worst case'' for $p=1$ and $N=16$) and thus negligible. Therefore, we compare the
first line of Eq.~(\ref{rousetrimer}) to the simulation data. Figure~\ref{autokorr} shows
that the approximate consideration of stiffness by Eq.~(\ref{rousetrimer}) reasonably accounts
for the suppression of correlations below the Rouse prediction (\ref{phirou.eq}).
Similar approaches are pursued in \cite{alga,hwr} and tested against experiments and 
simulations \cite{rmaacfcf,spmr}.

Figure~\ref{rousekorr} tests another prediction of Eq.~(\ref{phirou.eq}), the orthogonality
of the Rouse modes, i.e., $\langle \bbox{X}_{p}(0)\bbox{X}_{q}(0)\rangle \sim \delta_{pq}$,
for two chain lengths $N=16$ and $N=128$. Whereas the auto-correlation of the first mode is 
1 to 2 orders of magnitude larger than cross-correlations with $p>1$ for both chain lengths,
this difference in magnitude is only preserved for the short chain when increasing $p$ towards
$p=N-1$ (see Fig.~\ref{rousekorr}b). On the other hand, cross-correlations of the same amplitude 
as the self-correlation develop between $p=127$ and the smallest modes for $N=128$. If one 
interpretes a difference of 1 to 2 orders of magnitude between self- and cross-correlations as 
a numerical realization of $\langle \bbox{X}_{p}(0)\bbox{X}_{q}(0)\rangle \sim \delta_{pq}$, the 
full spectrum of Rouse modes is only delta-correlated for $N=16$, while small and large modes 
interfere for large chain lengths.

\section{Dynamic properties of the melt}
\label{dyn}
This section discusses the simulation results for dynamic properties of the polymer melt. It
is split into three subsections. The first deals with an analysis of various mean-square
displacements. The presentation of the data is very much influenced by a recent thorough 
theoretical and numerical study of the Evans-Edwards model \cite{t,u1,u,sbe}. The second subsection
presents the relaxation behavior of the correlation function of the bond-vector, of the end-to-end 
distance and of the Rouse modes, whereas the last subsection discusses the chain-length dependence of
the corresponding relaxation times and of the diffusion coefficient.

\subsection{Mean-Square Displacements}
\label{msd}
Important quantities to study the crossover from Rouse to reptation dynamics are the 
following mean-square displacements:
\begin{equation}
\begin{array}{lcl}
g_1(t) &=& \Big\langle \Big[\bbox{R}_{N/2}(t)-\bbox{R}_{N/2}(0)\Big]^2 \Big\rangle\;, \\[3mm]
g_2(t) &=& \Big\langle \Big[\bbox{R}_{N/2}(t)-\bbox{R}_{N/2}(0)-\bbox{R}_{\rm cm}(t)+
\bbox{R}_{\rm cm}(0) \Big]^2 \Big\rangle \;,\\[3mm]
g_3(t) &=& \Big\langle \Big[\bbox{R}_{\rm cm}(t)-\bbox{R}_{\rm cm}(0)\Big]^2 \Big\rangle \;,\\[3mm]
g_4(t) &=& \frac{\displaystyle 1}{\displaystyle 2} \Big (\Big\langle \Big[\bbox{R}_1(t)-\bbox{R}_1(0)\Big]^2 
\Big\rangle + \Big\langle \Big[\bbox{R}_N(t)-\bbox{R}_N(0)\Big]^2 \Big\rangle\Big) \;.
\end{array}
\label{eq:msd}
\end{equation}
Here, $g_1$ is the mean-square displacement of the inner monomer of a chain (situated at
position $\bbox{R}_{N/2}(t)$ at time $t$), $g_2$ is the same displacement measured relative to 
the motion of the chain's center of mass ($\bbox{R}_{\text{cm}}(t)$ is position of the center
of mass at time $t$), $g_3(t)$ is the mean-square displacement of the center of mass, and 
$g_4$ is that of the end monomers. Within the framework of reptation theory one expects \cite{b,a,d,o}
\begin{equation}
g_1(t) \sim \left\{
\begin{array}{lll}
t^{1/2}     &\mbox{\,if}& t \ll \tau_{\text{e}}\;, \\
t^{1/4}     &\mbox{\,if}& \tau_{\text{e}} \ll t \ll \tau_{\text{R}}\;,\\
(t/N)^{1/2} &\mbox{\,if}& \tau_{\text{R}} \ll t \ll \tau_{\text{d}}\;,\\
t/N^2       &\mbox{\,if}& \tau_{\text{d}} \ll t\;,
\end{array}
\right .
\qquad
g_3(t) \sim \left\{
\begin{array}{lll}
t/N           &\mbox{\,if}& t \ll \tau_{\text{e}} \;,\\
(t/N^2)^{1/2} &\mbox{\,if}& \tau_{\text{e}} \ll t \ll \tau_{\text{R}}\;,\\
t/N^2         &\mbox{\,if}& \tau_{\text{R}} \ll t\;.
\end{array}
\right .
\label{g1g3rep}
\end{equation}
At early times, a chain does not feel the entanglement constraints imposed by its neighbors. 
Assuming the Rouse model to provide a realistic description for non-entangled chains, one expects 
$g_1\sim t^{1/2}$ and $g_3 \sim t$ in this case. If $t$ equals the entanglement time 
$\tau_{\text{e}}$, the constraints begin to dominate the polymer dynamics. For larger times, a 
chain moves as if it were confined in a tube created by its neighbors: It can only slide along 
the tube axis (the so-called ``primitive path''), whereas motion perpendicular to it is 
suppressed. Since the tube can be thought of as an envelope around the chain's random-walk-like 
configuration, the chain performs Rouse dynamics along a random walk. This leads to $g_1\sim t^{1/4}$
and $g_3 \sim t^{1/2}$. If $t$ reaches the Rouse 
time $\tau_{\text{R}}$, the chain configuration is relaxed inside the tube. For larger times, the 
center of mass diffuses freely and the inner monomer begins to diffuse out of the tube. The
monomer leaves the tube completely if $t$ equals the disentanglement time $\tau_{\text{d}}$.

The power laws of Eq.~(\ref{g1g3rep}) represent the asymptotic behavior. They can be observed 
clearly only for very large $N$, where the time scales $\tau_{\text{e}}$, $\tau_{\text{R}}$ and 
$\tau_{\text{d}}$ are well separated. For smaller $N$ significant corrections must be expected.
This has recently been demonstrated by a detailed theoretical analysis \cite{u1} of the 
Evans-Edwards model \cite{ee} and by comparing the analytical results with the outcome of 
simulations \cite{t,u}. The Evans-Edwards model is a lattice model, in which a single 
chain moves on a simple cubic lattice through impenetrable obstacles placed in the centers of 
each unit cell of the lattice (a monomer occupies one lattice site). These obstacles impose a
strong geometric confinement on the motion of the chain: The tube diameter $d_{\text{T}}$ is of
the order of the lattice constant. Even under these very favorable conditions long chains and
long times are required to display the asymptotic power laws clearly \cite{t,u}. If one wants
to analyze the motion of the inner monomer in a different model, the experience with the studies 
of \cite{t,u1,u,sbe} suggests to use the following three quantities: the mean-square 
displacement $g_2$ which behaves as
\begin{equation}
g_2(t) \left \{
\begin{array}{lll}
\sim g_1(t) & \; \mbox{if} \; &t \ll \tau_{\text{d}}\;,\\
&&\\
\rightarrow R^2_{\text{g}} & \; \mbox{if} \; &\tau_{\text{d}} \ll t \; ,
\end{array}
\right.
\label{g2rep}
\end{equation}
so that a transition from $g_2 \sim t^{1/4}$ to $g_2 \sim t^{1/2}$ cannot be interpreted as a
crossover to free diffusion due to the saturation of $g_2$ at late times, the ``cubic 
invariant'' \cite{t,u1,u,sbe}
\begin{equation}
g_6(t) = \sqrt{\Big \langle \sum_{\alpha =1}^3 \Big[R_{\alpha,N/2}(t)-R_{\alpha,N/2}(0)
\Big]^4 \Big \rangle } \sim g_1(t) \;,
\label{defci}
\end{equation}
which is predicted to be much less plagued than $g_1$ by preasymptotic corrections (for the 
Evans-Edwards model), and the ratio $g_4(t)/g_1(t)$. This ratio should start around 1 at 
early times, develop a maximum at intermediate times and approach 1 if $t\rightarrow \infty$
(= free diffusive limit for both $g_4$ and $g_1$). For Rouse dynamics the maximum occurs
at $t \lesssim \tau_{\text{R}}$ and has the value $g_4/g_1=2$. If the chain reptates, the analysis
of the Evans-Edwards model suggests that the position of the maximum is still given by $t 
\approx \tau_{\text{R}}$, but its amplitude should be much larger than 2 ($g_4/g_1 \approx 4\sqrt{2}$ for
$N\rightarrow \infty$) \cite{u1,u,sbe}.

Figures~\ref{g6t4} and \ref{g2t4} depict the time dependence of $g_6$ and $g_2$,
respectively. The ordinates are divided by $t^{1/4}$ to highlight the expected asymptote
for the onset of reptation. The figures show that the initial increase of $g_6$ and $g_2$, which 
is roughly compatible with a $t^{1/2}$-behavior for $200 \lesssim t \lesssim 5000$, considerably 
slows down when $t > 10^5$. However, an indication of the $t^{1/4}$-power law is only 
visible for $N=512$. Even for this chain length there is no clear sign of a subsequent $t^{1/2}$-increase,
but only a fairly protracted transition to the large-time limits $g_6 \sim t$ and $g_2 \rightarrow 
R^2_{\text{g}}$. All other chain lengths already cross over to these limits 
for $t > 10^5$ (this is especially well visible for $g_2$). Therefore, much larger 
(or stiffer \cite{mwb,faller,fph}) chains and still better statistics are needed to clearly separate the
different power laws and to eventually distinguish between reptation theory and alternative
approaches, such as polymer mode-coupling theory \cite{ken1,ken2,schsz,i,fs_1997,fs_1997_jcp}. 

Nonetheless, estimates of the tube diameter $d_{\text{T}}$ and the entanglement time 
$\tau_{\text{e}}$ can be obtained from the figures by posing
\begin{equation}
g_6(\tau_{\text{e}}) \simeq g_2(\tau_{\text{e}}) \simeq g_1(\tau_{\text{e}}) = 
\frac{d^2_{\text{T}}}{3} \;,
\label{defdT}
\end{equation}
where $\tau_{\text{e}}$ is defined by the intersection point of the $t^{1/2}$- and $t^{1/4}$-power
laws. The factor $1/3$ in Eq.~(\ref{defdT}) was proposed in \cite{kg,pkg} and justified by the following
argument: The tube diameter can be interpreted as the end-to-end distance of a chain segment with 
$N_{\text{e}}$ monomers. This is the largest segment which does not feel the entanglement 
constraint yet. It relaxes as a Rouse chain. So, $\tau_{\text{e}}=\tau_{\text{R}}(N_{\text{e}})$
which gives $g_1(\tau_{\text{e}}) \approx d^2_{\text{T}}/3$ \cite{a}. Qualitatively, the factor
$1/3$ also appears reasonable. If we assume that the end-to-end vector $\vec{d}^2_{\text{T}}$ is
predominantly oriented along the tube axis, the average extension of the tube perpendicular to its 
axis should be proportional to the radius of gyration $d^2_{\text{T}}/6$ and thus smaller than 
$d^2_{\text{T}}$ \cite{sbe}. 

If one accepts these arguments, Eq.~(\ref{defdT}) yields $d_{\text{T}} \approx 17.9, 19.3$ and 
$\tau_{\text{e}} \approx 84053, 100175$ for $g_2$ and $g_6$, respectively, so that on average 
$d_{\text{T}} \approx 18.6$ and $\tau_{\text{e}} \approx 92114$. Using the result for $d_{\text{T}}$ 
we can furthermore estimate the entanglement length $N_{\text{e}}$ by requiring $d_{\text{T}}^2=
R^2_{\text{e}}(N_{\text{e}})$. To this end, we fitted the simulation data for $R^2_{\text{e}}$
in the range $16 \leq N \leq 64$ by a power law, which gives $R^2_{\text{e}}(N)=6.872N^{1.08}$ so 
that $N_{\text{e}} \approx 38$. This value is about $26\%$ larger than the original estimate
\cite{m}, but coincides very well with the result of a recent study \cite{mwb}.

{ 
Figure~\ref{g4g1} shows the simulation results for the ratio of the mean-square displacement of the
end monomer and the inner monomer, $g_4(t)/g_1(t)$. The ratio starts around 1.5 at early times, 
exhibits a maximum at intermediate times and tends to 1 for large $t$. For $N\leq 64$ the maximum 
occurs close to $g_4/g_1 = 2$, i.e., to the value expected from the Rouse model, whereas it is larger 
for $N=128$ ($g_4/g_1 \approx 3$).
An increase of the maximum beyond the Rouse limit with growing chain length is predicted theoretically for 
the Evans-Edwards model and proposed as a sensible indicator of reptation dynamics \cite{u1,u,sbe}. For a
reptating chain the disparity between $g_4$ and $g_1$ has to increase because the motion of the
inner monomer is strongly confined by the tube at intermediate times, whereas end monomers always take 
part in tube renewal. Our data for $N=128$ are indicative of such an enhancement beyond the Rouse limit
and are qualitatively similar to simulation results of the Evans-Edwards model with a randomly
dilute obstacle lattice (see Fig.~5 of \cite{sbe}). Unfortunately, we have to restrict the discussion 
to $N \leq 128$ because the calculation of the ratio makes the statistical inaccuracies become so pronounced 
that a quantitative analysis for $N=512$ was not possible.
}

In addition to the displacement of inner and end monomers the motion of the center of mass is also
interesting. Figure~\ref{g3t34} plots $g_3(t)/t^{3/4}$ versus time for all chain lengths studied. 
This representation of the ordinate was motivated by Eq.~(\ref{g1g3rep}). If a chain reptates,
$g_3(t)/t^{3/4}$ should exhibit a minimum close to $t=\tau_{\text{R}}$, which is preceded by a
$t^{-1/4}$-descent and followed by a $t^{1/4}$-ascent. In analogy to $g_2$ and $g_6$, a 
signature of this behavior emerges for $N=512$ only, whereas the transition to free diffusion
intervenes for smaller chain lengths.

The data reveal that $g_3$ is subdiffusive at early times.
Particularly for $N=512$, where $\tau_{\text{e}}$ and the crossover-time to free diffusion are 
well separated, this behavior is very pronounced. The mean-square displacement increases as $g_3 
\sim t^x$ with an effective exponent $x \approx 0.84$. Subdiffusive motion of the center of mass 
at early times is not unusual. It is generally found in computer simulations of linear polymers
(see \cite{d,o} for reviews and \cite{psy} for a chemically realistic model of polyethylene) and of
rings \cite{mwc}, and it has also been observed lately in experiments (on short chains) \cite{spmr}. 
The subdiffusive behavior should be considered 
as a condensed-phase effect. In a dense melt, a polymer is intermingled with many other chains. 
Their presence impedes the motion of the center of mass of the tagged chain, leading to 
subdiffusive behavior. Evidence for this interpretation is provided, for instance, by simulations
at different densities. If the density decreases towards the single-chain limit, the 
bond-fluctuation model yields $g_3 \sim t$ for all times (i.e., typical Rouse behavior)
\cite{o,m,mwb}. Another evidence stems from recent molecular dynamics simulations of the Kremer-Grest
model \cite{y}. In this study, the velocity-autocorrelation function of the center of mass was
determined. The correlation function becomes negative at intermediate times and approaches 0
from below. This means that the surrounding polymers exert a force on the tagged chain, which 
reverses its velocity and tries to push it back to its original position.

Since the subdiffusive motion of $g_3$ is already present for short non-entangled chains, it should
replace the expectation from the Rouse model $g_3 \sim t$ for $t < \tau_{\text{e}}$ [see Eq.~(\ref{g1g3rep})].
Therefore, one can try to determine $\tau_{\text{e}}$ from the intersection of $g_3 \sim t^{0.84}$ and $g_3 
\sim t^{1/2}$. This yields $\tau_{\text{e}} \approx 268718$, which is about a factor of $2.5$ larger than 
the results derived from $g_6$ or $g_2$. Nevertheless, if one accepts this value, another estimate of $d_{\text{T}}$ 
can be obtained by posing 
\begin{equation}
g_3(\tau_{\text{e}}) = \frac{b^2 d_{\text{T}}^2}{3R_{\text{e}}^2} 
\left (\frac{k_{\text{B}}T}{\zeta b^2}\tau_{\text{e}}\right)^{1/2} \;,
\label{g3taue}
\end{equation}
where the factor $1/3$ stems from Eq.~(\ref{defdT}) and the requirement $g_1(\tau_{\text{d}})=
g_3(\tau_{\text{d}})$. In Eq.~(\ref{g3taue}), $\zeta$ denotes the monomeric friction coefficient and 
$\tau_{\text{e}}$ is (assumed to be) given by \cite{o}
\begin{equation}
\tau_{\text{e}}= \left(\frac{d_{\text{T}}}{b}\right)^4 \frac{\zeta b^2}{k_{\text{B}}T} \; ,
\label{taue_doi_ed}
\end{equation}
so that 
\begin{equation}
d_{\text{T}} = \Big [3g_3(\tau_{\text{e}}) R_{\text{e}}^2\Big ]^{1/4} \;.
\label{dT_from_g3taue}
\end{equation}
Using $g_3(\tau_{\text{e}})\approx 10.37$ and $R_{\text{e}}^2 = 5348$, Eq.~(\ref{dT_from_g3taue}) leads to 
$d_{\text{T}} \approx 20.3$. This value is compatible with previous estimates from $g_6$ and
$g_2$. However, due the small exponent $1/4$ in Eq.~(\ref{dT_from_g3taue}) the result for the tube
diameter is not very sensible to variations of $\tau_{\text{e}}$. For instance, when using the average
$\tau_{\text{e}}$ from $g_6$ and $g_2$, i.e., $\tau_{\text{e}}\approx 92114$, so that $g_3(\tau_{\text{e}})\approx 
4.27$, one obtains $d_{\text{T}} \approx 16.2$ which is also in reasonable agreement with the results derived from 
the monomer displacements. 

On the other hand, the power law $g_3 \sim t^{1/2}$ provides a further possibility to determine $d_{\text{T}}$
by the intersection with the diffusive behavior $g_3 \sim t$. The intersection point should define the Rouse time
$\tau_{\text{R}}$ ($\approx 4.34 \times 10^6$), and one expects the value $g_3(\tau_{\text{R}})$ to be close to 
$d_{\text{T}}^2$. If we determine the prefactor by the same argument which led to Eq.~(\ref{g3taue}), we obtain
$g_3(\tau_{\text{R}})=d_{\text{T}}^2/3$. Using $g_3(\tau_{\text{R}})\approx 54.8$, we find $d_{\text{T}} \approx 13$.
This value seems too small in comparison with the previous results which indicates that the Rouse time is 
presumably larger than determined. Therefore, $\tau_{\text{R}}=4.34 \times 10^6$ rather corresponds to a lower bound 
than to a reliable estimate.

\subsection{Correlation Functions}
\label{corfun}
If a chain starts to reptate, the motion of the monomers perpendicular to the primitive path is
restricted to displacements smaller than the tube diameter. This implies that the decay of 
orientational correlation functions of chain segments should significantly slow down if the size 
of the segment becomes comparable to $d_{\text{T}}$. Therefore, correlation functions which 
probe reorientations on different length scales along the backbone of a chain are interesting 
quantities.

The smallest segment along the backbone of a chain is the bond vector. The
correlation function of the bond vector can be defined by
\begin{equation}
\phi_{\text{b}}(t)=\frac{1}{(N-1)l^2}\sum_{n=1}^{N-1}\Big \langle \bbox{l}_n(t) \cdot
\bbox{l}_n(0) \Big \rangle \; ,
\label{defphib}
\end{equation}
where $l^2=\langle \bbox{l}^2(0)\rangle$ is the mean-square bond length. This quantity is also
interesting because it should be closely related to the shear modulus \cite{fs_1997_jcp}. 
Figure~\ref{phib} illustrates the time- and chain-length dependences of $\phi_{\text{b}}(t)$. The 
data are presented in a scaling plot where the time axis is given in units of the relaxation time,
$\tau_{\text{b}}$, of the bond vector defined by $\phi_{\text{b}}(\tau_{\text{b}})=1/\text{e}$.
This time scale depends weakly on chain length (see Table~\ref{tab1}). If $\tau_{\text{b}}$ 
was the only relevant time scale, all $\phi_{\text{b}}(t)$'s, measured for different $N$, should
collapse onto a master curve. Figure~\ref{phib} shows that such a collapse is realized
for about $75\%$ of the relaxation, whereas the final decay to zero depends on chain length.
Larger chains relax more slowly than shorter ones.

Such a behavior is already expected from the (discrete \cite{verdier}) Rouse model which expresses 
$\phi_{\text{b}}(t)$ as
\begin{equation}
\phi_{\text{b}}(t) = \frac{1}{N-1}\sum_{p=1}^{N-1} \phi_p(t) = \frac{1}{N-1}\sum_{p=1}^{N-1} 
\exp\Big(-\frac{tp^2}{\tau_{\text{R}}}\Big) \; ,
\label{eq:phibrouse}
\end{equation}
where $\tau_{\text{R}}$ is the Rouse time and $\phi_p(t)$ the correlation function of the 
$p$th Rouse mode \cite{a,verdier}
\begin{equation}
\phi_p(t)=\frac{\langle\bbox{X}_p(t)\cdot \bbox{X}_p(0)\rangle}{\langle\bbox{X}_p^2(0)\rangle}
=\exp\Big(-\frac{tp^2}{\tau_{\text{R}}}\Big) \; .
\label{eq:phiprouse}
\end{equation}
If $t \ll
\tau_{\text{R}}$, the relaxation behavior of the sum in Eq.~(\ref{eq:phibrouse}) is dominated
by large $p$'s, which yields $\phi_{\text{b}}(t) \sim 1/t^{1/2}$ independent of $N$ (in the 
continuum limit). On the other hand, if $t \gtrsim \tau_{\text{R}}$, $\phi_{\text{b}}(t)$ 
relaxes the more slowly, the longer the chain length. These predictions seem to be in accord with 
the simulation results. However, there are differences at short times for all chain lengths and 
at long times for large $N$. 

At short times, the Rouse formula (\ref{eq:phibrouse}) underestimates the relaxation strength of 
$\phi_{\text{b}}(t)$. This discrepancy is caused by Eq.~(\ref{eq:phiprouse}) which assumes an
exponential decay for all Rouse modes. However, especially the simulation results for large Rouse 
modes do not decay in a simple exponential fashion (see discussion of Fig.~\ref{rousekorr16}).
These large modes determine the behavior of $\phi_{\text{b}}(t)$ at short times, whereas the 
late-time relaxation is dominated by the smallest Rouse mode which is almost exponential for $N 
\leq 128$. Nevertheless, Eqs.~(\ref{eq:phibrouse}) and (\ref{eq:phiprouse}) only provide a 
reasonable description for the $N$-dependent tail of $\phi_{\text{b}}(t)$ if $N=16$.
For larger $N$ the simulation data relax more slowly and finally develop a power-law decay if
$N\geq 128$. The power law starts around $t \approx \tau_{\text{e}}$ and extends over about 3 
decades up to the end of our simulation.  With the present simulation results we cannot 
decide whether the power law is an indication of a possible two-step relaxation which would become clearly 
visible for longer chains, as suggested by polymer mode-coupling theory \cite{i,fs_1997_jcp}.

Qualitatively, the slow power-law decay of $\phi_{\text{b}}(t)$ implies that some bonds lose the 
memory of their original orientation only very gradually. Persistence of bond orientation is also 
an important notion 
in the development of the continuum theory of reptation \cite{a}. There, a key quantity is the 
probability ${\psi}(s,t)$ that a segment $s$ of the primitive chain is still in the tube at time $t$. 
This probability is related to the projection of the (unit) tangent vectors, $\bbox{u}(s^\prime,t)$, of all 
segments $s^\prime$ onto one primitive path segment $s$, i.e., onto $\bbox{u}(s,0)$ (see 
Eq.~(6.45) of Ref.~\cite{a}). In analogy to ${\psi}(s,t)$, we define
\begin{equation}
\phi_n(t)=\frac{\sum^N_{n'=1}\langle \bbox{l}_{n'}(t)\cdot \bbox{l}_n(0)\rangle}
{\sum^N_{n'=1}\langle \bbox{l}_{n'}(0)\cdot \bbox{l}_n(0)\rangle} \; ,
\label{tubelife.eq}
\end{equation}
which represents the probability that the original orientation of the bond vector $\bbox{l}_n(0)$ 
is still present at time $t$. This quantity can be calculated by the (discrete \cite{verdier}) Rouse
model. The result is
\begin{eqnarray}
\phi_n(t)
&=&
\frac{\sum_{p=1 \,(\text{odd})}^{N-1}\big[ \sin(np\pi/N)/\tan(p\pi/2N)\big]\exp\big(-p^2t/\tau_{
\text{R}}\big)}{\sum_{p=1\,(\text{odd})}^{N-1}\sin(np\pi/N)/\tan(p\pi/2N)}
\label{tlrouse.dis} \\
&\stackrel{p/N\ll 1}{=}&
\frac{4}{\pi}\sum_{p=1 \atop (p\, \text{odd})}^\infty\frac{1}{p}\sin(np\pi/N)\exp\big(-p^2t/\tau_{\text{R}}
\big) \label{tlrouse.con} \;.
\end{eqnarray}
The continuum limit (\ref{tlrouse.con}) is identical to the reptation
formula for ${\psi}(s,t)$ if $\tau_{\text{R}}$ is replaced by the disentanglement
time $\tau_{\text{d}}$. Numerically, the discrete and continuum results for $\phi_n(t)$ are
almost indistinguishable. Therefore, we compare Eq.~(\ref{tlrouse.dis}) with simulation data for 
$N=16$ and $N=128$ in Fig.~\ref{bij16+128}. 

If $N=16$, Eq.~(\ref{tlrouse.dis}) provides a fairly accurate description of the simulated
$\phi_n(t)$. Both theory and simulation yield bell-shaped curves whose amplitudes become smaller
as time increases. This behavior can be rationalized as follows: Since chain ends are more mobile 
than inner monomers, bond vectors close to the ends decorrelate more quickly than those in the 
inner part of the chain. The decorrelation thus propagates from the chain ends towards the middle 
monomer. The bond vector orientation of the inner monomer is rather long-lived. Even at $t=\tau_
{\text{ee}}$, where the end-to-end vector correlation has decayed to about $30\%$ of its original
value (see Eq.~(\ref{deftee}) and Fig.~\ref{recorr}), $\phi_n(t)$ is still about 0.5 for $n=8$. 
This slow relaxation is already present for short, Rouse-like chains and should not be considered 
as a characteristic feature of reptation-like dynamics, since the Rouse and reptation formulas agree 
with one another [Eq.~(\ref{tlrouse.con})]. 

However, the simulation results for $\phi_n(t)$ exhibit differences between short and long chains. 
As Fig.~\ref{bij16+128} shows, the relaxation of inner monomers becomes strongly retarded compared to 
the theoretical (reptation) prediction if $N=128$. More than $50\%$ of the initial orientation of the 
inner third of the chain is preserved even at $\tau_{\text{ee}}$. These long-lived correlations, which 
are not present for small $N$, should be responsible for the power-law behavior of $\phi_{\text{b}}(t)$ (see 
Fig.~\ref{phib}) because they begin to develop in the same time window as the power law, i.e, for 
times larger than the entanglement time ($\tau_{\text{e}}/\tau_{\text{ee}} \approx 0.04$ for 
$N=128$), and last until $t > \tau_{\text{ee}}$ (the time $t/\tau_{\text{b}} = 10^4$ in Fig.~\ref{phib}
corresponds to about $t/\tau_{\text{ee}}\approx 2$ in Fig.~\ref{bij16+128}b). 

Furthermore, they should also become visible in the motion 
of the monomers, particularly when comparing the mean-square
displacement of the end monomer, $g_4(t)$, with that of 
the inner monomer $g_1(t)$. Figure~\ref{bij16+128} suggests
that the ratio $g_4(t)/g_1(t)$ for $t \approx \tau_{\text{ee}}$
is much larger for $N=128$ than for $N=16$. Our simulation
data for $g_4/g_1$ support this expectation (see Fig.~\ref{g4g1}).

If the chain length increases, one may expect the slowly relaxing zone around the middle monomer 
to grow and to influence the dynamic behavior of the chain more and more. For instance, one 
might speculate that the end-to-end vector relaxes in two steps: There is a first step initiated
by the chain ends. Their fast motion decorrelates adjacent bond vectors very efficiently.
However, this decorrelation does not propagate homogeneously along the polymer backbone towards
the inner monomer as it is the case for short chains, but slows down in the inner part of the chain. The
relaxation of this part is responsible for the second step. 

Some evidence for this conjecture is given in Fig.~\ref{recorr} which shows the time-dependence
of the end-to-end vector correlation function, $\phi_{\text{e}}(t)$, for all chain lengths 
studied and compares the simulation data with the Rouse (or reptation) prediction
\begin{equation}
\phi_{\text{e}}(t)=\frac{\big \langle \bbox{R}_{\text{e}}(t) \cdot \bbox{R}_{\text{e}}(0)
\big \rangle}{\big \langle \bbox{R}_{\text{e}} (0)^2 \big \rangle}
=\sum_{p=1\atop (p\, \text{odd})}^\infty \frac{8}{p^2\pi ^2} \exp(-p^2t/\tau_{\text{ee}})\;,
\label{phirecorr.eq}
\end{equation}
where the relaxation time $\tau_{\text{ee}}$ is defined by
\beq
\phi_{\rm e}(\tau_{\rm ee})= \sum_{p=1\atop (p\, \text{odd})}^\infty \frac{8}{p^2 \pi^2}\exp(-p^2)
=0.298221 \;.
\label{deftee}
\eeq
Note that Eq.~(\ref{phirecorr.eq}) is the same for both Rouse and reptation theory, the only
difference being that $\tau_{\text{ee}}$ is the Rouse time, $\tau_{\text{R}}$, in the first case 
and the disentanglement time, $\tau_{\text{d}}$, in the latter case, i.e.,
\begin{equation}
\tau_{\text{ee}} = \left \{
\begin{array}{lll}
\tau_{\text{R}} = \frac{\displaystyle 1}{\displaystyle 3\pi^2}\, \frac{\displaystyle \zeta b^2}{\displaystyle 
k_{\rm B}T}\, N^2& \;\mbox{if}\; & N \ll N_{\text{e}} \;,\\
&&\\
\tau_{\text{d}} = \frac{\displaystyle 1}{\displaystyle \pi^2}\, \frac{\displaystyle \zeta b^2}{\displaystyle 
k_{\rm B}T}\, \frac{\displaystyle b^2}{\displaystyle d^2_{\text{T}}}\, N^3 & \;\mbox{if} \; & N \gg N_{\text{e}} \; .
\label{taurd.eq}
\end{array}
\right.
\eeq
where prefactors $1/3\pi^2$ and $1/\pi^2$ are taken from \cite{a} (they have been set equal to 1 in the 
discussion of the mean-square displacements in Sect.~\ref{msd}).

Equation~(\ref{phirecorr.eq}) suggests that simulation data 
for $\phi_{\text{e}}(t)$ depend on $N$ only via $\tau_{
\text{ee}}$. By using a rescaled time axis, $t/
\tau_{\text{ee}}$, it should be possible to 
superimpose all data onto a master curve which is given by 
Eq.~(\ref{phirecorr.eq}). Figure~\ref{recorr} confirms this
expectation if $N\leq 128$. For $N=512$, however, the 
relaxation is more complicated. It seems to occur in two steps. 
The first step starts at about $t \approx \tau_{\text{e}}$, and 
the second takes place in the time window where $g_3$ crosses over 
to free diffusion, i.e., on the scale of the Rouse time 
$\tau_{\text{R}}$ (see Fig.~\ref{g3t34}). This behavior is 
similar to that expected from polymer mode-coupling theory
\cite{i,fs_1997,fs_1997_jcp}. The theory predicts two power laws:
$1-\phi_{\text{e}}(t) \sim t^{9/32}$ for $\tau_{\text{e}} \ll t \ll 
\tau_{\text{R}}$ and $1-\phi_{\text{e}}(t) \sim t^{3/8}$ for $\tau_{\text{R}} 
\ll t \ll \tau_{\text{RR}}$ ($\tau_{\text{RR}}=$ renormalized Rouse
time $ < \tau_{\text{d}}$). Our data are compatible with a $t^{9/32}$-behavior
for about one decade ($0.002 \lesssim t/\tau_{\text{ee}} \lesssim 0.1$) and 
eventually with $t^{3/8}$ close to $\tau_{\text{ee}}$. Of course, this can
only be considered as an indication that these predictions might be relevant
for our model. Clarification of this point requires a much better separation
of the relevant time scales and thus simulations of longer chains. 
On the other
hand, it is not obvious that Eqs.~(\ref{phirecorr.eq}) and (\ref{deftee}) are
accurate for chain lengths in the crossover regime from Rouse to reptation
behavior, although they describe both limiting cases of the simple Rouse
and the ``pure'' (i.e., asymptotic) reptation theory. Certainly, a quantitatively
reliable theoretical prediction for $\phi_{\text{e}}(t)$ as well as for $\tau_{\text{ee}}$
in the regime where $N$ and $N_{\text{e}}$ are comparable would be very desirable.

The observed two-step-like relaxation behavior is not limited to $\phi_{\text{e}}(t)$. It
can also be observed for the correlation function of
the Rouse modes $\phi_p(t)$. Figure~\ref{rousekorr16} shows
scaling plots of $\phi_p(t)$ for $N=16$ and $N=512$. As
suggested by Eq.~(\ref{eq:phiprouse}), the scaling time,
$\tau_p$, is defined by $\phi_p(\tau_p)=1/\text{e}$. The
figure illustrates that the first Rouse mode is almost
exponential for $N=16$, whereas the relaxation of higher 
modes becomes progressively non-exponential with increasing 
$p$ (smaller distance along the chain backbone). However,
the curves cannot be described (completely) by a stretched 
exponential. This would imply that the fast decay of, say,
$p=15$ for $\phi_{\text{e}}(t) > 1/\text{e}$ entails a
correspondingly slower relaxation if $\phi_{\text{e}}(t) < 
1/\text{e}$. Since the data splay out with increasing $p$ at
short times, they should also splay out at late times, but
in reverse order. There is no indication of that behavior
in Fig.~\ref{rousekorr16}a. The interplay of chain stiffness
and local excluded volume forces gives rise to a more
complicated than stretched-exponential time dependence. The
same behavior is also observed for a bead-spring model of
short chains \cite{bbpb}, and, to some extent, for $N=512$ 
in Fig.~\ref{rousekorr16}b. For large chains, however,
there is another factor which may contribute: cross-correlations 
between Rouse modes. Figure~\ref{autokorr} revealed that
these can be as pronounced as the self-correlation if $p=
N-1$. Since $\phi_{p=1}(t)$ exhibits again the signature of a
two-step relaxation, such as $\phi_{\text{e}}(t)$, this behavior
could also influence the decay of $\phi_{p=511}(t)$.

\subsection{Correlation Times and Diffusion Coefficient}
\label{cordif}
The discrete Rouse model yields the following expression 
for the relaxation time $\tau_p$ \cite{verdier}
\begin{eqnarray}
\tau_p
&=&
\frac{1}{12}\,\frac{\zeta}{k_{\text{B}}T} \,
\left[\frac{b}{\sin(p\pi/2N)}\right]^2 
\label{eq:taup1} \\
&=&
\frac{2}{3}\,\frac{N\zeta}{k_{\text{B}}T} \,
\big\langle \bbox{X}_p^2(0)\big \rangle \;,
\label{eq:taup2}
\end{eqnarray}
where Eq.~(\ref{phirou.eq}) was used to obtain the last line. The
unit of $\tau_p$ is set by the monomeric friction coefficient
$\zeta$ which should be independent of the mode index $p$ and chain
length $N$. Within the framework of the Rouse model $\zeta$
can be determined from the diffusion coefficient, $D$, of a 
chain by
\begin{equation}
ND = \frac{k_{\text{B}}T}{\zeta} \; .
\label{defdiffrouse}
\end{equation}
If one furthermore identifies the effective bond length with
$b^2=R^2_{\text{e}}/(N-1)$ (as already done in Fig.~\ref{autokorr}), 
Eq.~(\ref{eq:taup1}) suggests to
construct a master curve by plotting $12DN(N-1)\tau_p/
R_{\text{e}}^2$ versus $p/N$. Figure~\ref{tp} shows that
such a scaling is only possible if $N \leq 64$. For these
chain lengths the scaling yields a curve which is independent of $N$
and coincides with the Rouse prediction for $p/N\lesssim
0.06$. On the other hand, there are deviations for $p/N \geq 0.3$,
where Eq.~(\ref{eq:taup1}) overestimates the simulation results.

Qualitatively, this behavior is identical to that observed for $\langle
\bbox{X}^2_p(0) \rangle$ in Fig.~\ref{autokorr}. This
suggests that the discrepancy between simulation and 
Eq.~(\ref{eq:taup1}) might be related to local stiffness effects and
that it could be removed by expressing $\langle \bbox{X}_p^2(0)
\rangle$ via Eq.~(\ref{rousetrimer}) or by inserting the simulation results 
for $\langle \bbox{X}_p^2(0)\rangle$ in  Eq.~(\ref{eq:taup2}). 
The latter choice was made in the inset of Fig.~\ref{tp}.
This yields a good description for $N \leq 64$. Two conclusions
can be drawn from that: First, as expected from Eq.~(\ref{eq:taup2}),
$\langle \bbox{X}_p^2(0)\rangle$ carries the whole $p$-dependence
so that $\zeta$ is constant. This should be considered as a property of
the bond-fluctuation model and not as a general feature because a
bead-spring model can exhibit a different behavior \cite{bbpb}%
\footnote{This point was not stated explicitly in Ref.~\cite{bbpb}.
However, since roughly $\langle \bbox{X}_p^2(0) \rangle \sim p^{-2.2}$ and 
$\tau_p \sim p^{-2}$ in this case, Eq.~(\ref{eq:taup2}) would imply
$\zeta = \zeta_p \sim p^{0.2}$. Similar observations were also made
in neutron scattering experiments on polyisobutylene \cite{rmaacfcf}. 
A quantitative description of the experimental results was only possible 
if both local stiffness and an increase of $\zeta$ with increasing mode index 
$p$ were taken into account.}.
Second, although the diffusion coefficients for $N\leq 64$ do not obey 
Eq.~(\ref{defdiffrouse}) [see Fig.~\ref{td+dn}], the Rouse relation (\ref{eq:taup2}) 
effectively remains valid. This is possible because the simulation results for
$\tau_p$ can be written as $\tau_p=f(N)[\sin(p\pi/2N)]^{-2}$ and the dependence of 
the function $f$ on $N$ is just given by $(N(N-1)D/R_{\text{e}}^2)^{-1}$.
Theoretically, $N(N-1)D/R_{\text{e}}^2$ should represent the reorientation time
of a monomer, $\zeta b^2/k_{\text{B}}T$, which is independent of $N$. In practice,
$N(N-1)D/R_{\text{e}}^2$ varies from about $6 \times 10^{-4}$ for $N=16$ to about 
$3.5 \times 10^{-4}$ for $N=64$, i.e., by about $40\%$. For larger chain length, however,
the residual $N$-dependence of $\tau_p$ can no longer be lumped in a prefactor,
although $N(N-1)D/R_{\text{e}}^2$ decreases to about $7 \times 10^{-5}$ for $N=512$
so that it is 1 order of magnitude smaller than for $N=16$. Nonetheless, $\tau_p$
depends more strongly on $N$ than $N(N-1)D/R_{\text{e}}^2$ for small $p$ and does not
seem to be any longer a function of $p/N$ alone. Even if one changes the prefactor to 
shift the data for $N=128$ onto those for $N=512$, one can guess from Fig.~\ref{tp} that 
such a collapse is only possible for $p/N \gtrsim 0.03$, whereas $\tau_p$ for $N=512$ becomes
larger than that for $N=128$ when $p/N < 0.03$. Since scaling with $p/N$ was possible for
$\langle \bbox{X}_p^2(0)\rangle$, which probes the static properties of a chain only,
Fig.~\ref{tp} suggests that the friction coefficient $\zeta$ becomes $p$- and $N$-dependent.

Finally, Fig.~\ref{td+dn} shows the variation of $\tau_{\text{ee}}$ [defined by Eq.~(\ref{deftee})] 
and of $D$ with chain length. The ordinates were scaled by the chain length dependence expected 
from the Rouse model, i.e., $\tau_{\text{ee}}/N^2$ and $ND$ [see Eqs.~(\ref{taurd.eq}) and 
(\ref{defdiffrouse})]. The figure illustrates that neither $\tau_{\text{ee}}$ nor $D$ behave 
Rouse-like for small $N$ or reptation-like for large $N$. A possible explanation for this finding
is given by a recent reanalysis \cite{mwb} of the crossover scaling from the semi-dilute-solution to 
the melt state of the bond-fluctuation model \cite{m}. This new analysis focused on the diffusion
coefficient, but substantially varied chain stiffness, which had not been done before. The 
important outcome for the present work is that the diffusion data of Fig.~\ref{td+dn} just lie in
the crossover regime between Rouse dynamics, which is realized by flexible chains in dilute solution,
and reptation dynamics which is most pronounced for stiff chains in the melt. Apparently, chain stiffness
reduces the tube diameter and thereby amplifies the reptation behavior. The same conclusion has also 
been suggested by a recent molecular-dynamics simulation of the Kremer-Grest model \cite{faller}. Since 
the diffusion data belong to the crossover regime, this should also hold for $\tau_{\text{ee}}$ in 
Fig.~\ref{td+dn} so that neither $\tau_{\text{ee}}\sim N^2$ nor $\tau_{\text{ee}}\sim N^3$ can be 
observed. If one just fitted the last two chain lengths ($N=128,512$) by a power law, one would obtain 
$\tau_{\text{ee}} \sim N^{3.23}$. This is similar to experiments where one often finds that the relaxation 
time scales as $N^{3.4}$ \cite{j,c}. Of course, the present simulation data cannot prove, but only provide 
an indication that the bond-fluctuation model approaches (the possible asymptotic behavior) $\tau_{\text{ee}} 
\sim N^3$ from below with an exponent larger than 3, as suggested theoretically \cite{a,k2,u1}. These 
theoretical studies show that the crossover to the reptation asymptote is postponed to very large $N$. To
observe this asymptote one has to study either much longer or (eventually) stiffer chains which seem to 
show a clearer signature of reptation for accessible chain lengths \cite{mwb,faller}. 

Nevertheless, one can use the simulation results for $\tau_{\text{ee}}$ to estimate the entanglement length
$N_{\text{e}}$ by requiring that $\tau_{\text{e}}$ corresponds to the relaxation time of a chain with
end-to-end vector $d^2_{\text{T}}=R^2_{\text{e}}(N_{\text{e}})$. To read off $N_{\text{e}}$, we fitted 
$\tau_{\text{ee}}$ in the range of $\tau_{\text{e}}$ ($=92114$), i.e., for $16 \leq N \leq 64$, by a power law. 
This gives $\tau_{\text{ee}}=24.71 N^{2.31}$ so that $N_{\text{e}} \approx 35$. This value is compatible with
our previous estimate from $d^2_{\text{T}}=R^2_{\text{e}}(N_{\text{e}})$,
i.e., $N_{\text{e}} \approx 38$ (see Sect.~\ref{msd}), so that
on average $N_{\text{e}} \approx 37$. This result agree very well 
with that obtained by the above mentioned reanalysis of the crossover scaling 
\cite{mwb}.

\section{Conclusions}
\label{conc}
The present paper summarizes an attempt to extend previous
work \cite{o,m,wittmer} on the dynamics of long chain 
polymer melts towards larger chain lengths and longer
times. The aim was to sufficiently enter the reptation
regime to perform a careful test of the theory. However,
it turned out that the crossover from non-entangled to
entangled dynamics is very gradual. Even the longest
chain, $N=512$, which is about 13 times larger
than the entanglement length $N_{\text{e}}$, can only be 
considered as slightly entangled when analyzed in terms of 
the asymptotic power laws of reptation theory. This finding
is expected from a theoretical analysis of the Evans-Edwards 
model \cite{t,u1,u,sbe} and molecular dynamics simulations
\cite{kg,pkg}. 

For $N=512$ we find evidence
that there is an entanglement time $\tau_{\text{e}}$, above 
which the mean-square displacements of inner monomers, 
$g_1(t)$ [$\approx g_6(t)$] and $g_2(t)$, and that of the 
center of mass, $g_3(t)$, increase as $\sim t^{1/4}$ and
$\sim t^{1/2}$, respectively. From these quantities the
tube diameter $d_{\text{T}}$ and $\tau_{\text{e}}$ can be 
estimated: $d_{\text{T}}\approx 18.6$ and $\tau_{\text{e}}
\approx 92114$. Using these values and requiring 
$R^2_{\text{e}}(N_{\text{e}})=d^2_{\text{T}}$ and 
$\tau_{\text{ee}}(N_{\text{e}})=\tau_{\text{e}}$, the 
entanglement length can be read off of the simulation data:
$N_{\text{e}} \approx 37$. The quoted values for $d_{\text{T}}$ 
and $\tau_{\text{e}}$ are averages derived from
$g_2$ and $g_6$. The results from $g_2$ and $g_6$ were
rather close, whereas those from $g_3$ sometimes
deviate substantially and thus appear less reliable 
(perhaps due to preasymptotic effects, as pointed out in 
\cite{sbe}). Of course, there are still uncertainties
connected with these numbers, especially with $d_{\text{T}}$,
since the prefactors are unknown and $N=512$ is not really 
asymptotic. Here, the important point 
is that they are consistent with one another and with other
independent studies. For instance, the value $N_{\text{e}} 
\approx 37$ agrees very well with that found in \cite{mwb}.

In addition to mean-square displacements evidence for
reptation-like dynamics can be found in orientational
correlation functions. Bond vectors in the innermost part 
of a chain decorrelate from their initial orientations more
slowly than bonds close to the ends. This difference is
present for all chain lengths, but becomes much more
pronounced if $N \gg N_{\text{e}}$ and $\tau_{\text{e}} \ll
t \ll \tau_{\text{ee}}$. If there was a confining tube around 
a chain, such a behavior would be expected because 
reorientations of inner monomers are strongly hindered as
long as the geometric constraints exist (i.e., for $t < 
\tau_{\text{ee}}$), whereas chain ends always take part in 
tube renewal and thus dissolve the confinement immediately.
The simulations suggest that sensible indicators of this
behavior are a late-time power law decay of the bond-vector
correlation function and a modulated (eventually two-step)
relaxation of the end-to-end vector or Rouse modes 
correlation functions as well as the ratio $g_4(t)/g_1(t)$,
as proposed in \cite{u1,u,sbe}. The challenging problem
consists in determining these functions with sufficient
statistical accuracy in the relevant range of chain lengths
and times.

For short chains ($N\lesssim N_{\text{e}}$) the Rouse
model represents a viable, though not perfect approach.
Deviations are observed particularly on short length
scales along the polymer backbone. This is to be expected
because the microscopic potentials of the polymer should 
strongly influence the local dynamics. In general, these potentials 
cannot be represented by a flexible concatenation of 
Rousian harmonic springs, but introduce some local 
stiffness. If stiffness is taken into account, one can
obtain a (almost) quantitative description of the 
static correlation $\langle \bbox{X}^2_p(0) \rangle$ and
of the $p$-dependence of the Rouse modes' correlation
time $\tau_p$, which scales as $\tau_p \sim \langle 
\bbox{X}^2_p(0) \rangle$. Whereas the former property 
could be a general feature, the scaling $\tau_p \sim 
\langle \bbox{X}^2_p(0) \rangle$ is certainly a property
of our model. Simulations of other models \cite{bbpb} and
neutron scattering experiments on polyisobutylene 
\cite{rmaacfcf} suggest that the monomeric friction
coefficient can be mode dependent. Such findings are
plausible because one would in general not expect that 
static properties can uniquely determine the dynamic 
behavior. 

Another characteristic dynamic property of the Rouse 
modes is the progressive non-exponential character of the 
relaxation with increasing mode index $p$. This behavior
is not a special feature of the bond-fluctuation model, but
is also observed for a bead-spring model \cite{bbpb} and
in chemical realistic simulations of polyethylene
\cite{psy}. These deviations from a pure exponential decay
are important if one wants to calculate the relaxation of
quantities, for which large modes and short times also contribute 
significantly. An example is provided by the bond-vector
correlation function (see Fig.~\ref{phib}). If one inserts
the simulated Rouse modes instead of Eq.~(\ref{eq:phiprouse})
into Eq.~(\ref{eq:phibrouse}), a perfect description of the correlation
function is obtained. Since this description only requires
the orthogonality of the Rouse modes, it shows that the
melt dynamics does not introduce (significant) correlations
between different Rouse modes, i.e., $\langle \bbox{X}_p(t)
\cdot \bbox{X}_q(0) \rangle \propto \delta_{pq}$. We believe that
this property is general, since it is also observed for other
models \cite{bbpb}.

{ 
A further important feature is that there is no significant regime of 
chain lengths where the Rouse prediction $D \sim 1/N$ holds (see inset
of Fig.~\ref{td+dn}). Furthermore, the related prediction $g_3(t) \sim t$
(for $t \gtrsim 1$ in our units) is also not valid, not even for
$N=16$ (see Fig.~\ref{g3t34}). If $g_3(t) \sim t$ were true, the log-log 
plot $g_3/ t^{3/4}$ versus $t$ should exhibit a straight line with slope $1/4$.
However, we find a curved behavior spread over several decades. Our results
(Fig.~\ref{td+dn}) imply that there are about two decades in chain length
($10 \leq N \leq 10^3$) which are neither fully described by the simple Rouse
model nor by asymptotic reptation theory. A more complete theoretical 
description is thus called for.
}

\section*{Acknowledgement}
We are indebted to J.-L. Barrat, R. Everaers, F. Eurich, M.
Fuchs, P. Maass, C. Mischler, W. Paul, J. Wittmer for helpful 
discussions on various aspects of this work. This study would not have been 
possible without a generous grant of simulation time by the HLRZ J\"ulich, the RHRK
Kaiserslautern and the computer center at the University of Mainz. Financial support 
by the ESF Programme on 
``Experimental and Theoretical Investigation of Complex Polymer Structures'' (SUPERNET) 
is gratefully acknowledged.

%
%
\begin{table}
\begin{tabular}{rrrcrrr}
$N$ & $R_{\text{e}}^2$ & $R_{\text{g}}^2$ & $\langle \lambda_1 \rangle : \langle \lambda_2 \rangle : 
\langle \lambda_3 \rangle $ &$\tau_{\text{b}}$ & $\tau_{\text{ee}}$ & $\tau_{\text{1}}$ \\
\hline
16  & 136  & 23  & $12.42 : 2.78 : 1$ & 391 & 15180     & 14519     \\
32  & 293  & 49  & $12.30 : 2.74 : 1$ & 454 & 78509     & 76879     \\
64  & 607  & 102 & $12.17 : 2.71 : 1$ & 430 & 367863    & 384993    \\
128 & 1314 & 217 & $11.88 : 2.67 : 1$ & 445 & 2387464   & 2558552   \\
512 & 5348 & 885 & $11.87 : 2.65 : 1$ & 461 & 214679565 & 233045839 \\
\end{tabular}

\medskip

\caption[]{Chain length dependence of the size of a chain and of its relaxation times: 
$R^2_{\text{e}}=$ mean-square end-to-end vector,
$R^2_{\text{g}}=$ mean-square radius of gyration,
$\langle \lambda_\alpha \rangle=$ mean eigenvalue of the gyration tensor in spatial direction
$\alpha$ ($=1,2,3$) [see Eq.~(\ref{eq:gtensor})],
$\tau_{\text{b}}=$ relaxation time of the bond vector defined by $\phi_{\text{b}}(\tau_{\text{b}})=
1/\text{e}$,
$\tau_{\text{ee}}=$ relaxation time of the end-to-end vector defined by Eq.~(\ref{deftee}),
$\tau_{1}=$ relaxation time of the first Rouse mode defined by $\phi_{\text{b}}(\tau_{\text{b}})=
1/\text{e}$.}
\label{tab1}
\end{table}
%
%
\begin{figure}
\caption[]{Mean-square end-to-end vector, $R_{\text{e}}^2$ (upper symbols), and radius of gyration, 
$R_{\text{g}}^2$ (lower symbols), plotted against chain length $N$. Asymptotically, both 
quantities are expected to behave as $R_{\text{e}}^2 = C_\infty l^2 (N-1)$ and $R_{\text{g}}^2 = 
C_\infty l^2 (N-1)/6$ for dense melts. Here, $l=\langle \bbox{l}^2\rangle^{1/2}$ ($=2.64$) is the mean 
bond length and $C_\infty$ the characteristic ratio at infinite chain length. The value of $C_\infty 
\simeq 1.56$ was obtained by an extrapolation of $R_{\text{e}}^2/(N-1)l^2$ to $N\rightarrow \infty$ 
using $N > 100$. Both $R_{\text{e}}^2$ and $R_{\text{g}}^2$ enter the asymptotic regime if $N>100$
(for $R_{\text{g}}^2$ this is hard to see on the scale of this figure; see Fig.~\ref{lambda}). 
Furthermore, the figure shows that the present data are consistent with previously obtained 
results (on smaller systems) \cite{m,wittmer}.} 
\label{rerg}
\end{figure}
\begin{figure}
\caption[]{Eigenvalues, $\langle \lambda_\alpha \rangle /(N-1)$ ($\alpha=1,2,3$), of the gyration tensor 
{\sf Q} [Eq.~(\ref{eq:gtensor})] plotted versus chain length $N$. If the shape of a polymer were
spherical, one would have $\langle \lambda_1\rangle=\langle \lambda_2\rangle=\langle \lambda_3\rangle$. 
The result $\langle \lambda_1\rangle\gg\langle \lambda_2\rangle> \langle \lambda_3\rangle$ shows that it 
rather resembles a flattened ellipsoid (``prolate'' chains). The ratio of the eigenvalues ($\langle \lambda_1
\rangle:\langle \lambda_2\rangle:\langle \lambda_3\rangle = 11.87:2.65:1$ for $N=512$) is fairly
close to that expected for random walks ($12.03:2.72:1$ \cite{jtc}). As anticipated from Eq.~(\ref{lambda.eq}),
the eigenvalues sum up to the radius of gyration calculated independently by $R^2_{\text{g}}=
\sum_{n=1}^N(\bbox{R}_n-\bbox{R}_{\text{cm}})^2/N$, where $\bbox{R}_n$ and $\bbox{R}_{\text{cm}}$ 
are the position vectors of the $n$th monomer and of the center of mass, respectively.}
\label{lambda}
\end{figure}
\begin{figure}
\caption[]{Rescaled static autocorrelation of the Rouse modes, $8N(N-1)\langle \bbox{X}_p^2(0)
\rangle/R^2_{\text{e}}$, versus $p/N$ for all chain lengths studied ($R_{\text{e}}=$ end-to-end
distance). This rescaling is motivated by Eq.~(\ref{phirou.eq}). The dashed and the solid lines 
shows the theoretical predictions for fully flexible chains [Eq.~(\ref{phirou.eq})] and for 
freely rotating chains [Eq.~(\ref{rousetrimer})] where local stiffness due to the bond angle is 
taken into account. If $p/N \lesssim 0.05$ (corresponding to subchains larger than about 20 
monomers), both predictions coincide and describe the simulation data well. In this case, 
Eqs.~(\ref{phirou.eq}) and (\ref{rousetrimer}) are well approximated by the power law $p^{-2}$
(indicated by a solid line in the figure). However, if $p/N>0.3$ 
(corresponding to subchains smaller than about 3 monomers), the fully flexible (standard) Rouse 
model overestimates the correlation, whereas the (approximate) consideration of chain stiffness 
still provides a reasonable description. A magnification of the comparison between 
Eqs.~(\ref{phirou.eq}) and (\ref{rousetrimer}) is shown in the inset for $N=512$.}
\label{autokorr}
\end{figure}
\begin{figure}
\caption[]{Test of the orthogonality of the Rouse modes at $t=0$ [see Eq.~(\ref{phirou.eq})]. 
Figure~(a) shows $\langle \bbox{X}_1(0) \cdot \bbox{X}_q(0)\rangle$ ($q=1,\ldots, 127$) for $N=128$ 
($> N_{\text{e}} \approx 37$; $N_{\text{e}}=$ entanglement length). The inset magnifies 
the results for $q=2,\ldots, 127$. Cross-correlations between the first and the $q=2,3$ modes 
are about 1 to 2 orders of magnitude smaller than $\langle \bbox{X}_1^2(0)\rangle$,
but much stronger than those for $q>3$. If $N < N_{\text{e}}$, the behavior is qualitatively the
same as shown in the inset. However, all correlations are more than 2 orders of magnitude smaller than
the self-correlation for non-entangled chains. Figure~(b) compares $\langle \bbox{X}_p(0)\cdot\bbox{X}_q(0)\rangle$ 
of the largest mode for $N=128$, $\langle \bbox{X}_{127}(0)\cdot\bbox{X}_q(0)\rangle$, with that for 
$N=16$ $(<N_{\text{e}})$, $\langle \bbox{X}_{15}(0)\cdot\bbox{X}_q(0)\rangle$. If $N > N_{\text{e}}$, 
cross-correlations with the smallest Rouse modes are as large as the self-correlation, whereas they 
are at least an order of magnitude smaller if $N < N_{\text{e}}$.}
\label{rousekorr}
\end{figure}
\begin{figure}
\caption[]{Time dependence of $g_6(t)/t^{1/4}$ for different chain lengths $N$. $g_6(t)$ is the cubic invariant
defined by Eq.~(\ref{defci}). This function is predicted to behave similarly to the mean-square displacement of
an inner monomer $g_1(t)$, but to exhibit the characteristics of reptation more clearly than $g_1(t)$ \cite{t,u1,u,sbe}.
In the interval $200 \lesssim t \lesssim 5000$, $g_6(t)$ increases roughly as $t^{1/2}$. Then, it crosses over
to a weaker time-dependence which is compatible with $g_6(t)\sim t^{1/4}$ for $4 \times 10^5 \lesssim t \lesssim
2\times 10^6$. The intersection between these two asymptotic power-laws provides an estimate for the entanglement 
time $\tau_{\text{e}} \approx 100175$ (solid vertical line). Using Eq.~(\ref{defdT}), $g_6(\tau_{\text{e}})/
\tau_{\text{e}}^{1/4} \simeq d^2_{\text{T}}/3\tau_{\text{e}}^{1/4} \approx 7$, the tube diameter is approximately
given by $d_{\text{T}} \approx 19.3$. Furthermore, the two dashed vertical lines indicate the values of the
relaxation time, $\tau_{\text{ee}}$, of the end-to-end vector, defined by Eq.~(\ref{deftee}), for $N=16$ and 
$N=512$.}
\label{g6t4}
\end{figure}
\begin{figure}
\caption[]{Time dependence of $g_2(t)/t^{1/4}$ for different chain lengths $N$. $g_2(t)$ is the displacement
of an inner monomer relative to the chain's center of mass [Eq.~(\ref{eq:msd})]. For times shorter than the 
Rouse, $\tau_{\text{R}}$, or disentanglement time, $\tau_{\text{d}}$, one expects $g_2(t)\sim g_1(t)$, 
whereas $g_2(t) \rightarrow R^2_{\text{g}}$ otherwise. Due to Eq.~(\ref{taurd.eq}) these times can be 
represented by the relaxation time, $\tau_{\text{ee}}$, of the end-to-end vector [defined by  Eq.~(\ref{deftee})]. 
The latter behavior is borne out for $t > \tau_{\text{ee}}$ (vertical dashed lines shown for $N=16$ and $512$ only). 
As in Fig.~\ref{g6t4} for $g_6$ ($\approx g_1$), there is an approximate $t^{1/2}$-increase for $200 \lesssim t 
\lesssim 5000$ which crosses over to a weaker time dependence. This weaker dependence is (perhaps) compatible
with $g_2 \sim t^{1/4}$ for $N=512$ over about half a decade ($5\times 10^5 \lesssim t \lesssim 10^6$).
Using this interval the intersection with the extrapolation of the $t^{1/2}$-behavior yields estimates for
the entanglement time, $\tau_{\text{e}}$, and the tube diameter, $d_{\text{T}}$: $\tau_{\text{e}}\approx 84053$
(solid vertical line), $d_{\text{T}} \approx 18.6$ [via Eq.~(\ref{defdT})].}
\label{g2t4}
\end{figure}
\begin{figure}
\caption[]{Ratio of the mean-square displacement of the end monomer, $g_4(t)$, and the inner monomer, $g_1(t)$, versus
logarithm of time for different chain lengths $N$ [see Eq.~(\ref{eq:msd})]. The dashed vertical lines indicate the
relaxation time, $\tau_{\text{ee}}$, of the end-to-end vector for $N=16$ and $N=128$ [$\tau_{\text{ee}}$ is
defined by  Eq.~(\ref{deftee}); see Table~\ref{tab1}]. The vertical dotted shows the value of the maximum, $g_4/g_1=2$, 
expected from the Rouse model. This maximum should occur for $t \lesssim \tau_{\text{R}}$. For a reptating chain, a
maximum is still expected to occur at $t \approx \tau_{\text{R}}$, but to be much larger than 2
\cite{u1,u,sbe}. The onset of this behavior is visible for $N=128$. In the limit $t\rightarrow 
\infty$, both displacements become diffusive. So, $g_4/g_1 \rightarrow 1$. Since $g_4 \sim g_1$ in general, the calculation
of the ratio $g_4/g_1$ eliminates the dominant variation with time and highlights the differences between both 
displacements. Thereby, statistical errors also become much better visible. For $N=512$, they are so pronounced that
a quantitative analysis was not possible.}
\label{g4g1}
\end{figure}
\begin{figure}
\caption[]{Time dependence of $g_3(t)/t^{3/4}$ for different chain lengths $N$. $g_3(t)$ is the mean-square
displacement of the chain's center of mass [Eq.~(\ref{eq:msd})]. Since reptation theory predicts a crossover
from $g_3 \sim t^{1/2}$ for $\tau_{\text{e}} \lesssim t \lesssim \tau_{\text{R}}$ to $g_3 \sim t$ for $t > 
\tau_{\text{R}}$, $g_3(t)/t^{3/4}$
should exhibit a minimum at $t \simeq \tau_{\text{R}}$ and increase as $t^{-1/4}$ and $t^{1/4}$ for 
$t< \tau_{\text{R}}$ and $t> \tau_{\text{R}}$, respectively. Qualitatively, such a behavior is observed for
$N=512$ only (solid lines with slope $1/2$ and $1$). The intersection point of both power laws yields an 
estimate for the Rouse time $\tau_{\text{R}} \approx 4.34\times 10^6$, whence $d_{\text{T}} \approx 13$ by
virtue of $g_3(\tau_{\text{R}})=d^2_{\text{T}}/3$. The minimum is preceeded by another power-law increase 
(empirically, $g_3 \sim 
t^{0.84}$) if $t \lesssim \tau_{\text{e}}$. For smaller chain lengths the power law is not as well pronounced. 
The intersection point of the power laws $g_3 \sim t^{0.84}$ and $g_3 \sim t^{1/2}$ yields another estimate
of the entanglement time, $\tau_{\text{e}} \approx 268718$, and of the tube diamater, $d_{\text{T}} \approx
20.3$ [see Eq.~(\ref{dT_from_g3taue})]. $g_3$ becomes diffusive in the range $t \approx \tau_{\text{ee}}$
[indicated by dashed vertical lines for $N=16$ and $512$; $\tau_{\text{ee}}= $ relaxation time of the end-to-end
vector, see Eq.~(\ref{deftee})].}
\label{g3t34}
\end{figure}
\begin{figure}
\caption[]{Scaling plot of the autocorrelation function, $\phi_{\text{b}}(t)$, of the bond vectors for
different chain lengths versus $t/\tau_{\text{b}}$. The relaxation time, $\tau_{\text{b}}$, was defined by
(see Table~\ref{tab1}): $\phi_{\text{b}}(\tau_{\text{b}})=1/\text{e}$ (dashed horizontal lines in the main
figure and the inset). This scaling collapses the correlators onto a master curve if $0.25 \lesssim 
\phi_{\text{b}}(t) \leq 1$. The final decay, $\phi_{\text{b}}(t) < 0.25$, depends on chain length. For $N=16$, 
it is well described by the Rouse model, i.e., by Eq.~(\ref{eq:phibrouse}) (dashed curve in the main figure), 
whereas Eq.~(\ref{eq:phibrouse}) decays faster than
the simulation data for $N \geq 32$. Especially, the largest chain lengths ($N=128,512$) exhibit a 
slow power-law decay (emprically, $\phi_{\text{b}}(t) \sim t^{-0.35}$), as illustrated in the inset. The inset 
also indicates the Rouse behavior, $\phi_{\text{b}}(t) \sim t^{-0.5}$, expected for $t < \tau_{\text{e}}$ 
($\tau_{\text{e}}=92114=$ average entanglement time of Figs.~\ref{g6t4} and \ref{g2t4}).}
\label{phib}
\end{figure}
\begin{figure}
\caption[]{Survival probability (\ref{tubelife.eq}), $\phi_n(t)$, of an initial bond vector $\bbox{l}_n(0)$ for 
$N=16$ [panel (a)] and $N=128$ [panel (b)]. $\phi_n(t)$ is plotted as a function of the bond number $n$. For 
example, $n=1$ corresponds to the bond vector $\bbox{l}_1(0)$ connecting the first and the second monomer. Time is 
measured in units of the relaxation time of the end-to-end vector $\tau_{\text{ee}}$ (see Table~\ref{tab1}).
The solid lines represent the prediction of Eq.~(\ref{tlrouse.dis}). To improve the statistics the simulation results
were averaged over both halves of the chain -- the chain is symmetric with respect to its ends.  Therefore, 
results from $n=1$ to $n=N/2$ are shown only. This was sufficient for the short chain, but not for $N=128$. In this 
case, the bonds $n$ and $n+1$ were lumped additionally and the resulting average was plotted at the (non-existing) 
bond number $n+1/2$.}
\label{bij16+128}
\end{figure}
\begin{figure}
\caption[]{Autocorrelation function of the end-end vector, $\phi_{\text{e}}(t)$, versus scaled time
$t/\tau_{\text{ee}}$ for all chain lengths studied. The scaling time, $\tau_{\text{ee}}$, is defined by 
Eq.~(\ref{deftee}): $\phi_{\text{e}}(\tau_{\text{ee}})=0.298221$ (dashed horizontal line). The vertical dashed
line indicate the value of the entanglement time $\tau_{\text{e}}$ ($=92114=$ average $\tau_{\text{ee}}$ from 
Figs.~\ref{g6t4} and \ref{g2t4}) and of the Rouse time $\tau_{\text{R}}(N=512)$ [$\approx 4.34 \times 10^6$]
determined from Fig.~\ref{g3t34}. Both times are divided by $\tau_{\text{ee}}(N=512)$ (see Table~\ref{tab1}). 
The solid line depicts the result of Eq.~(\ref{phirecorr.eq}). The inset shows $1-\phi_{\text{e}}(t)$ versus
$t/\tau_{\text{ee}}$ for $N=512$ together with the power laws $1-\phi_{\text{e}}(t) \sim t^{9/32}$ and 
$1-\phi_{\text{e}}(t) \sim t^{3/8}$ expected from polymer mode-coupling theory for $\tau_{\text{e}} \ll t \ll 
\tau_{\text{R}}$ and $\tau_{\text{R}} \ll t \ll \tau_{\text{RR}}$ ($\tau_{\text{RR}}=$ renormalized Rouse time) 
\cite{i}.}
\label{recorr}
\end{figure}
\begin{figure}
\caption[]{Correlation function of the Rouse modes, $\phi_p(t)$, versus rescaled time 
$t/\tau_p$ for $N=16$ [panel (a)] et $N=512$ [panel (b)]. The scaling time, $\tau_p$, is defined by: $\phi_p
(\tau_p)= 1/\text{e}$ (dashed horizontal lines). For both chain lengths the relaxation of four representative mode 
indices $p$ ($=1,\ldots, N-1$) is shown. In addition, an exponential function (solid lines) is depicted. This
is the behavior expected from the Rouse model [see Eq.~(\ref{eq:phiprouse})]. The vertical dashed line in
panel (b) indicates the Rouse time $\tau_{\text{R}}(N=512)$ [$\approx 4.34 \times 10^6$] determined from 
Fig.~\ref{g3t34}. $\tau_{\text{R}}$ is divided by the relaxation time, $\tau_1$, of the first Rouse mode
(see Table~\ref{tab1}).}
\label{rousekorr16}
\end{figure}
\begin{figure}
\caption[]{Variation of the Rouse mode relaxation time with the mode index $p$ for different chain lengths 
$N$ (entanglement length $N_{\text{e}} \approx 37$). The relaxation time, $\tau_p$, is defined by the time
value at which the correlation function of the Rouse modes has decayed to $1/\text{e}$, i.e., $\phi_p(\tau_p)=
1/\text{e}$. Motivated by Rouse theory the abscissa and ordinate are scaled according to Eq.~(\ref{eq:taup1}) 
where $\zeta b^2/k_{\text{B}}T$ is identified with $R_{\text{e}}^2/N(N-1)D$ ($D=$ diffusion coefficient, see 
Eq.~(\ref{defdiffrouse}); $R_{\text{e}}^2=$ mean-square end-to-end vector).
This scaling leads to a reasonable collapse if $N \leq 64$, whereas the $p$-dependence for $N=128,512$
($\gg N_{\text{e}}$) is s-shaped. $\tau_p$ is larger than the Rouse prediction (solid line) if $p/N$ is small
(especially for $N=512$), but smaller than it for $p/N \rightarrow 1$. As for the static correlation
$\langle \bbox{X}^2(0) \rangle$ (see Fig.~\ref{autokorr}), the Rouse model provides a good description of
the $p$-dependence if $p/N \lesssim 0.05$ (and $N \leq 64$), but overestimates the relaxation time for $p/N>0.3$.
The inset therefore tests whether $\tau_p$ can be written as $\tau_p = 2\langle \bbox{X}^2(0) \rangle/3D$
[see Eq.~(\ref{eq:taup2})] with $\langle \bbox{X}^2(0) \rangle$ taken from the simulation (see Fig.~\ref{autokorr}). 
This is possible for $N \leq 64$, but not for large (entangled) chains, such as $N=512$.}
\label{tp}
\end{figure}
\begin{figure}
\caption[]{Chain length dependence of the end-to-end vector relaxation time, $\tau_{\text{ee}}$ (main figure), 
and of the diffusion coefficient, $D$, of a chain (inset). $\tau_{\text{ee}}$ and $D$ are determined from 
Eq.~(\ref{deftee}) and from the long-time limit of $g_3$ ($=\lim_{t\rightarrow \infty} g_3(t)/6t$), respectively. 
Both quantities are scaled by the $N$-dependences of Eqs.~(\ref{taurd.eq}) and (\ref{defdiffrouse}) to highlight 
deviations from Rouse behavior. The diffusion coefficient is compared with results from previous studies 
\cite{m,wittmer}. The solid lines in the main figure and
the inset indicate the behavior, $\tau_{\text{ee}} \sim 
N^3$ and $D \sim N^{-2}$, suggested by reptation theory. 
The triangles ($\triangle$) are the predictions of the Rouse formula $\tau_{\text{ee}}=(Nb)^2\zeta/3\pi^2k_{\text{B}}T=
R^2_{\text{e}}/3\pi^2 D$, where the simulation results are used for $D$ and $R^2_{\text{e}}$ (= end-to-end 
distance).}
\label{td+dn}
\end{figure}
%
%
\newpage
\setcounter{figure}{0}
\begin{figure}
\epsfysize=110mm
\epsffile{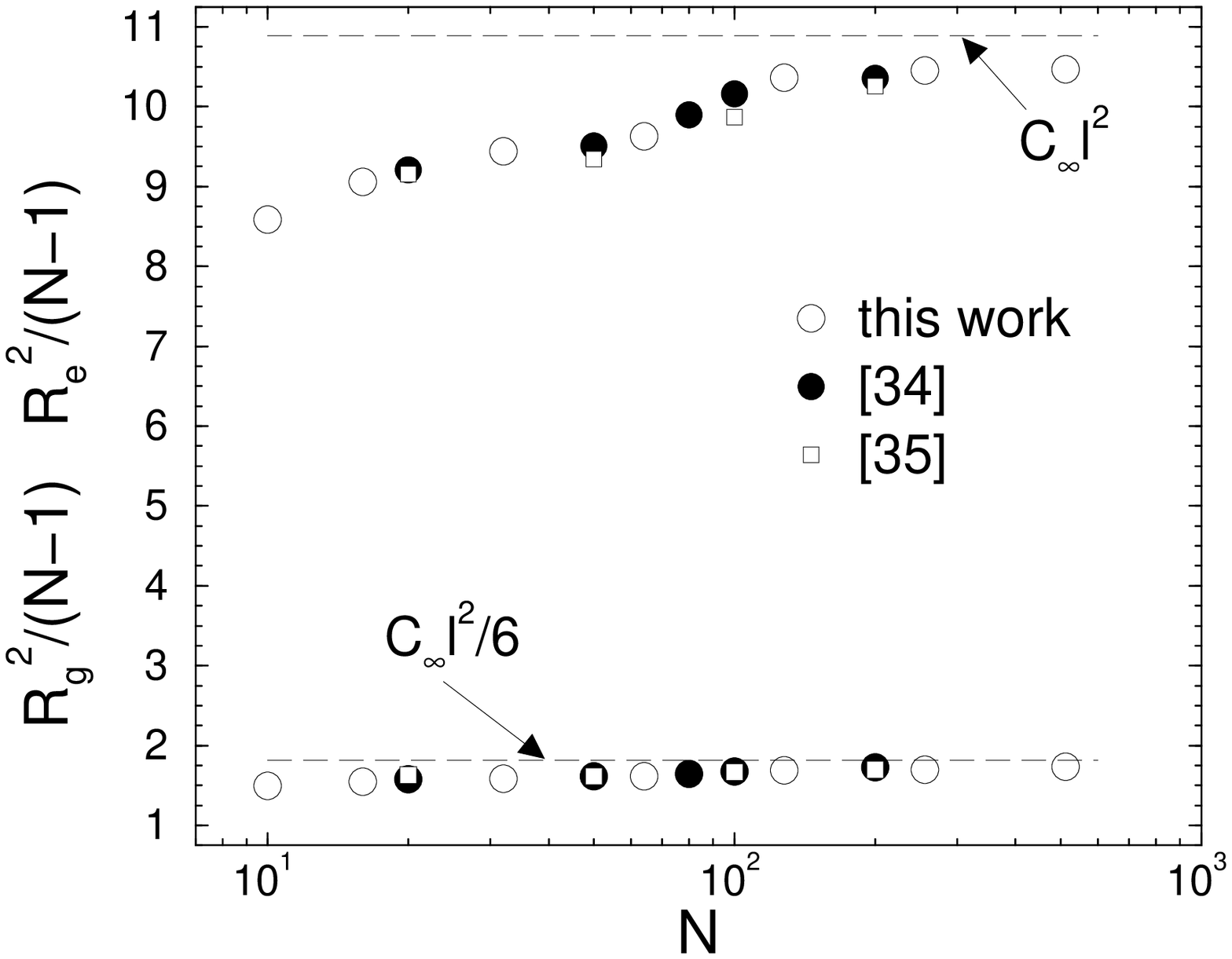}
\caption[]{}
\end{figure}     
\begin{figure}
\epsfysize=110mm
\epsffile{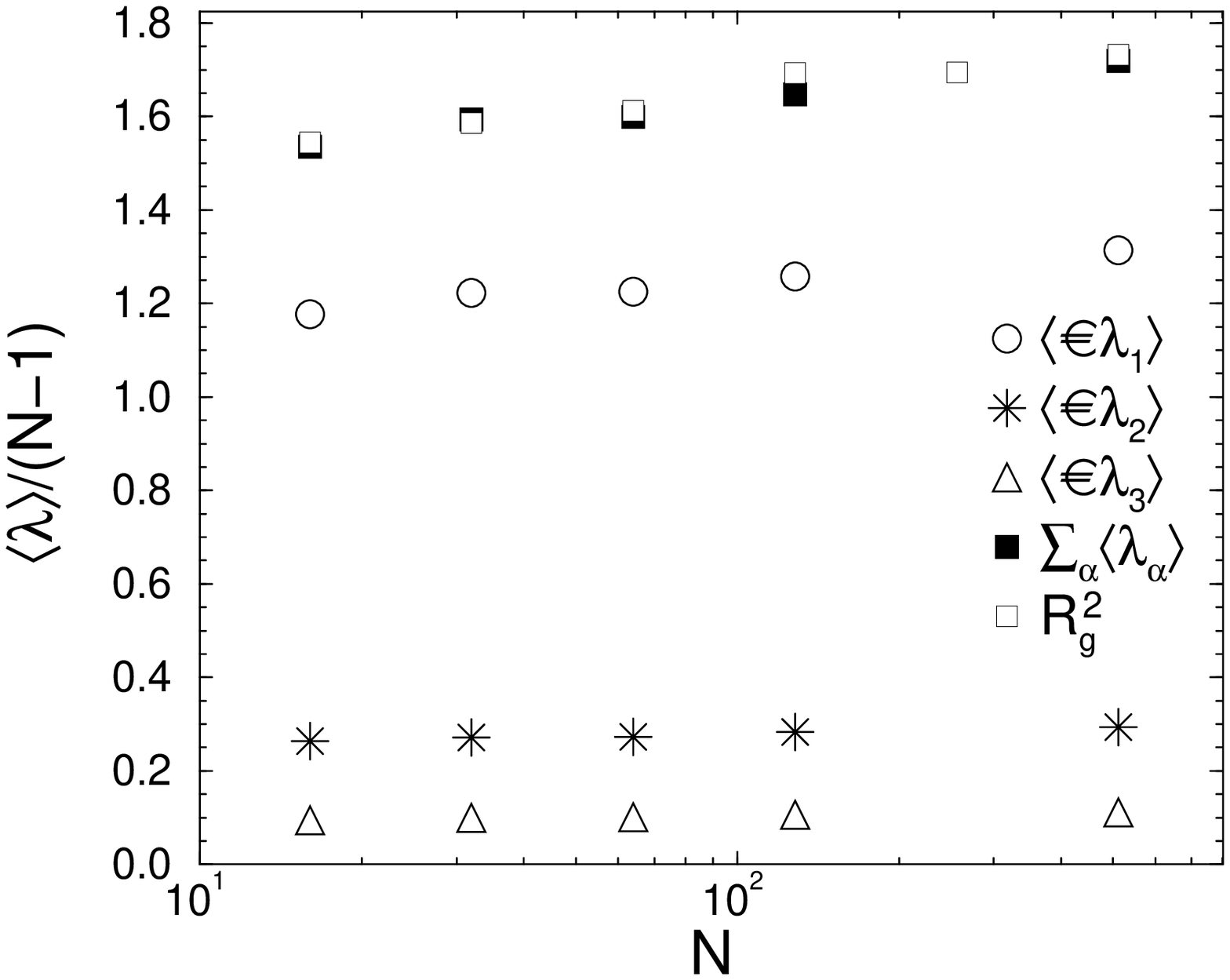}
\caption[]{}
\end{figure}     
\begin{figure}
\epsfysize=110mm
\epsffile{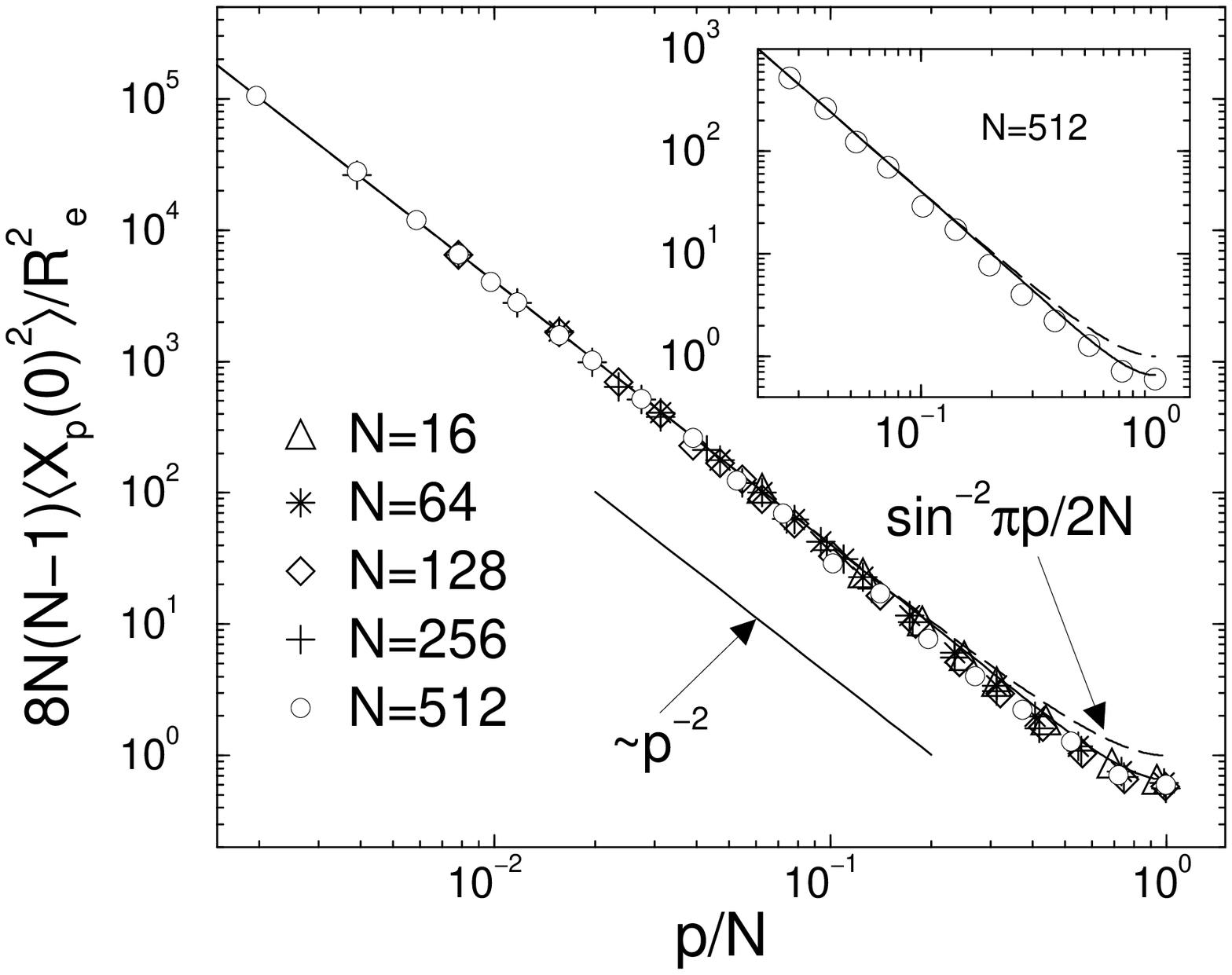}
\caption[]{}
\end{figure}     
\begin{figure}
\epsfysize=110mm
\epsffile{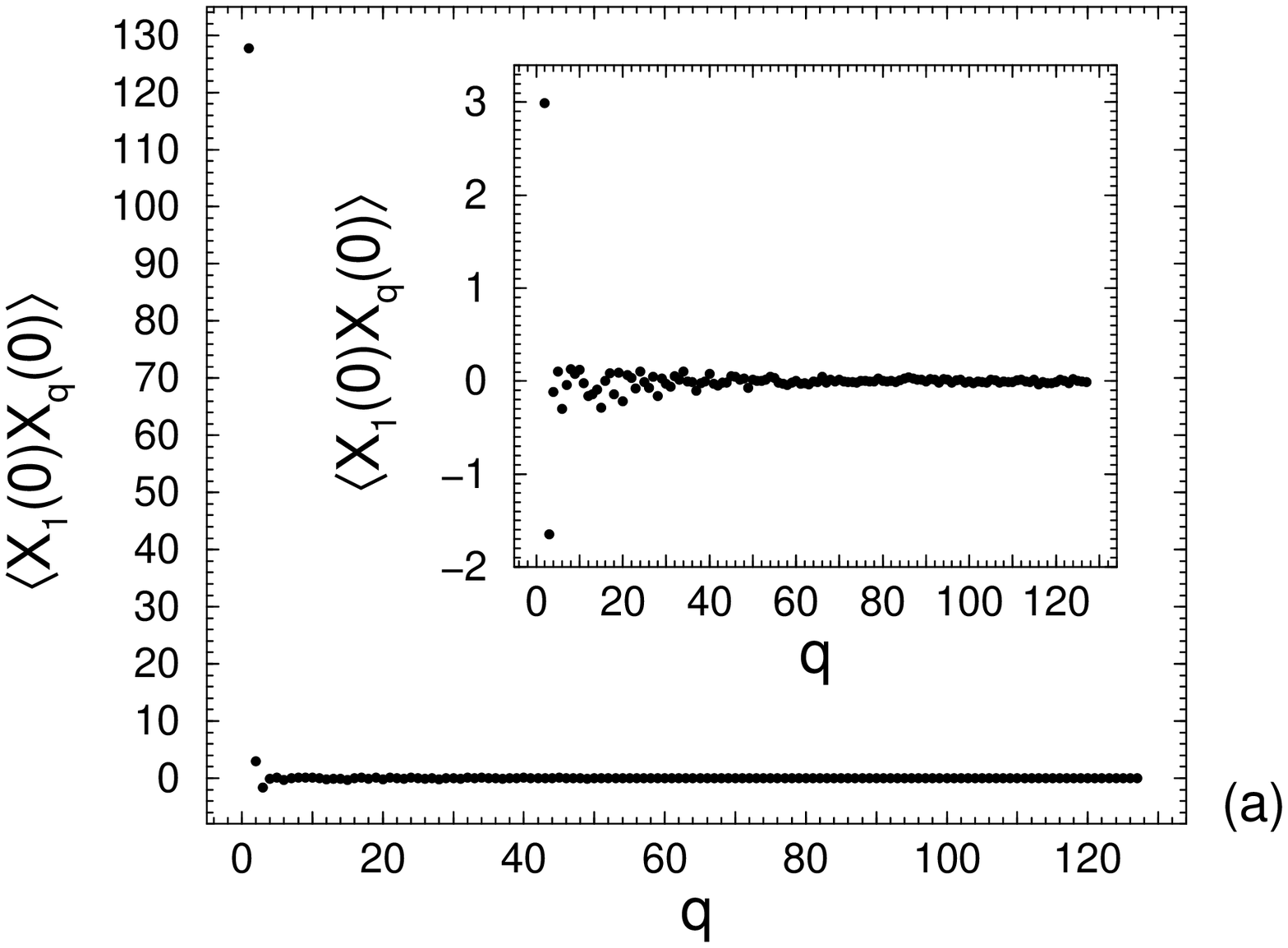}
\epsfysize=110mm
\epsffile{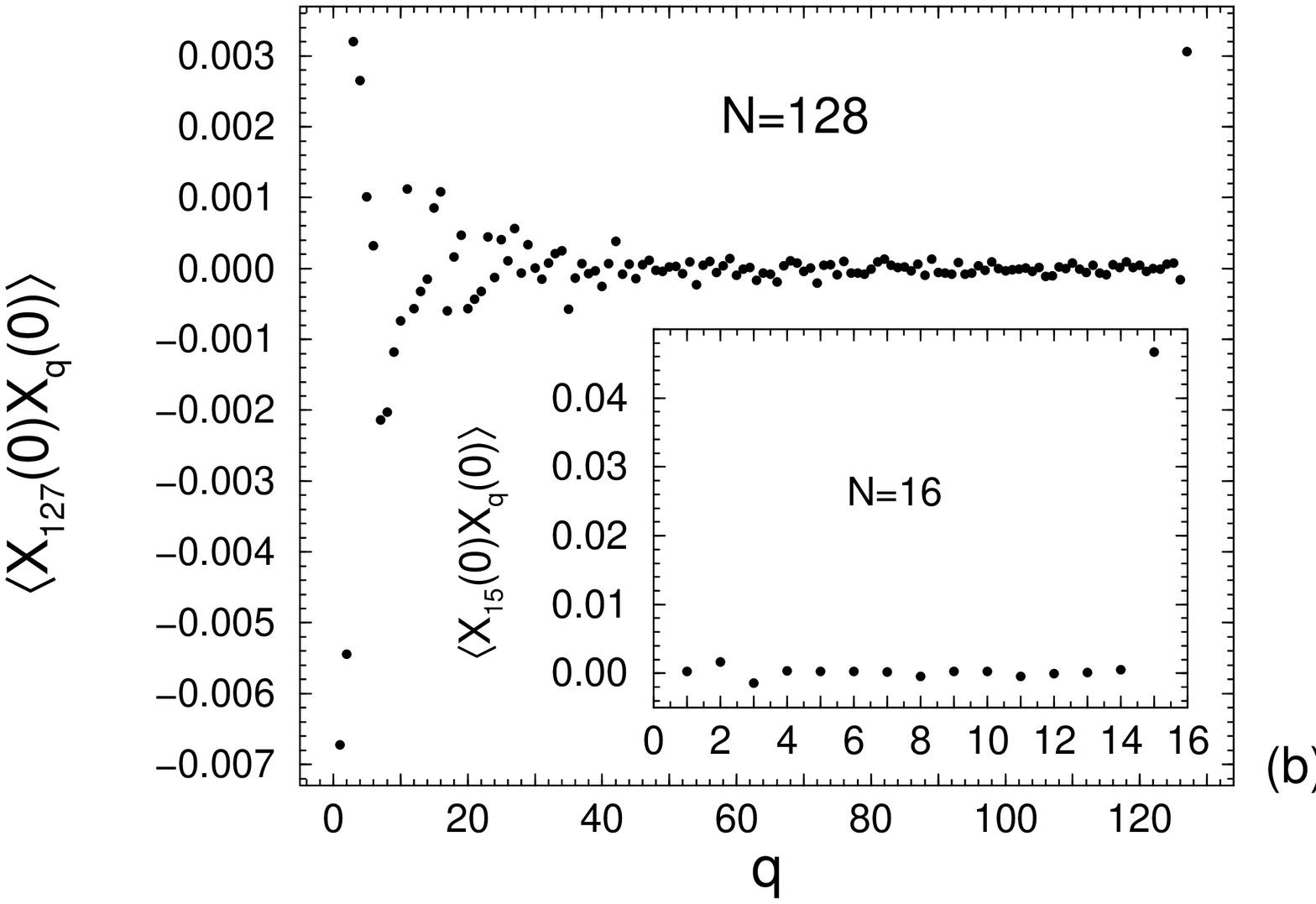}
\caption[]{}
\end{figure}     
\begin{figure}
\epsfysize=110mm
\epsffile{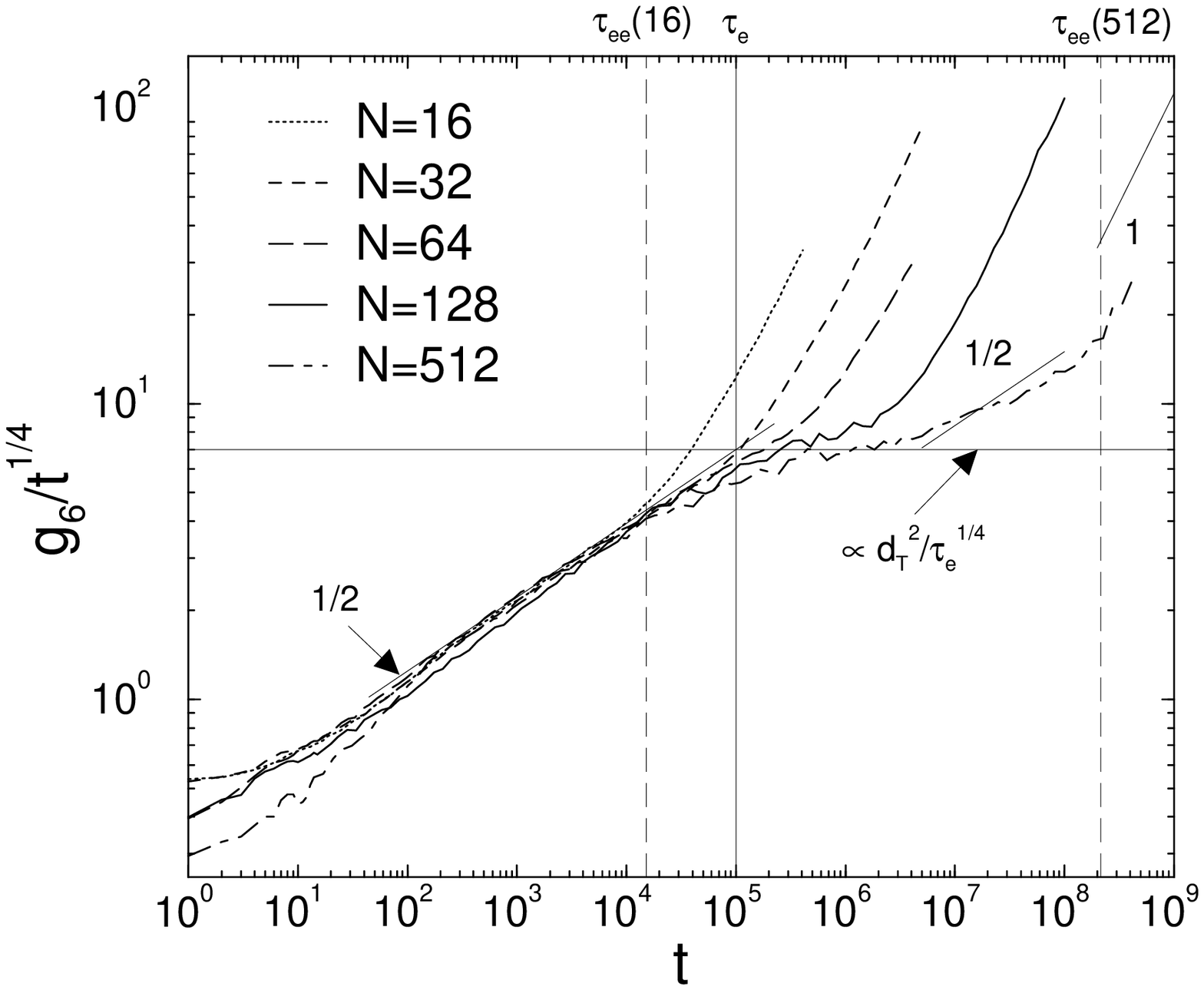}
\caption[]{}
\end{figure}
\begin{figure}
\epsfysize=110mm
\epsffile{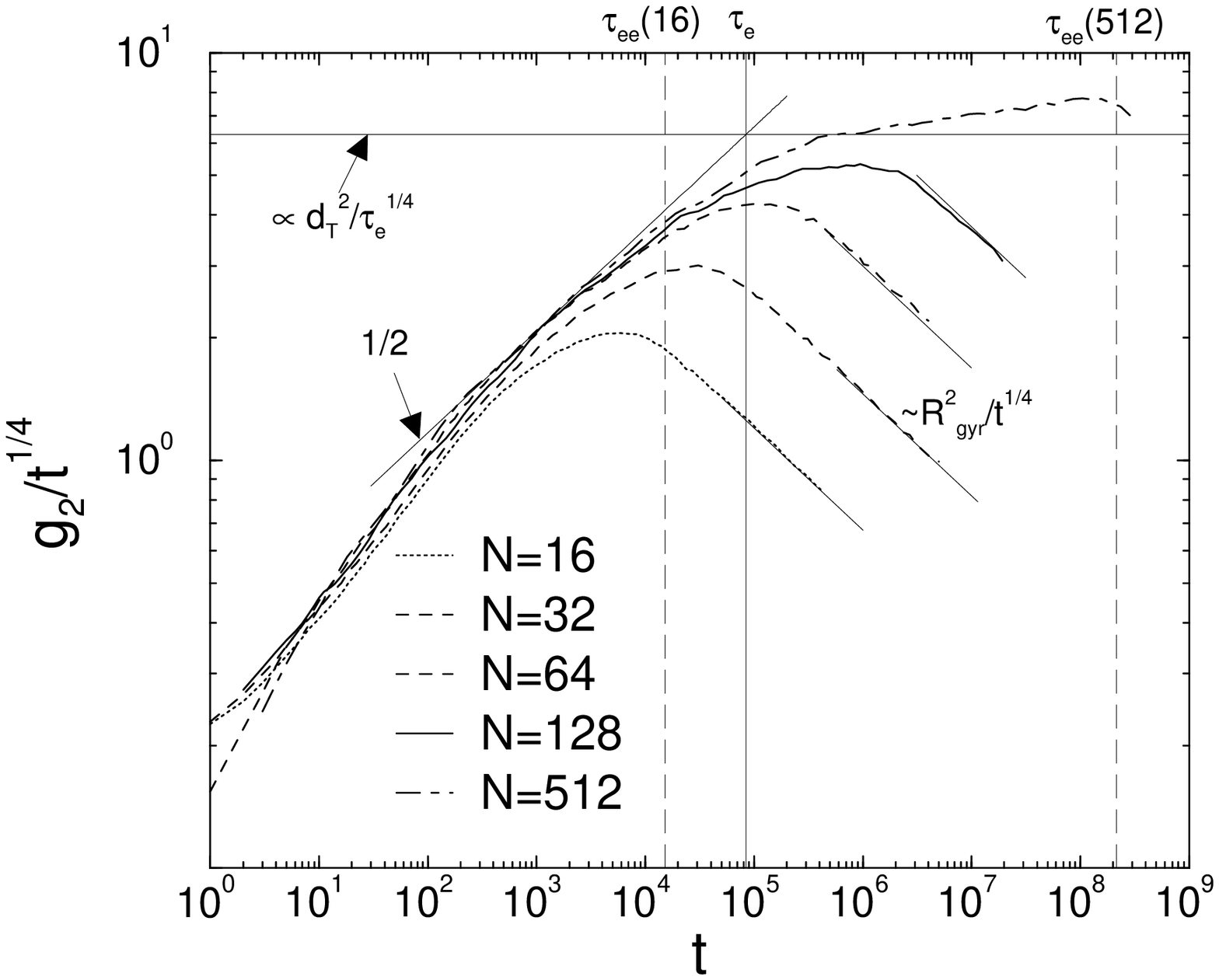}
\caption[]{}
\end{figure}
\begin{figure}
\epsfysize=110mm
\epsffile{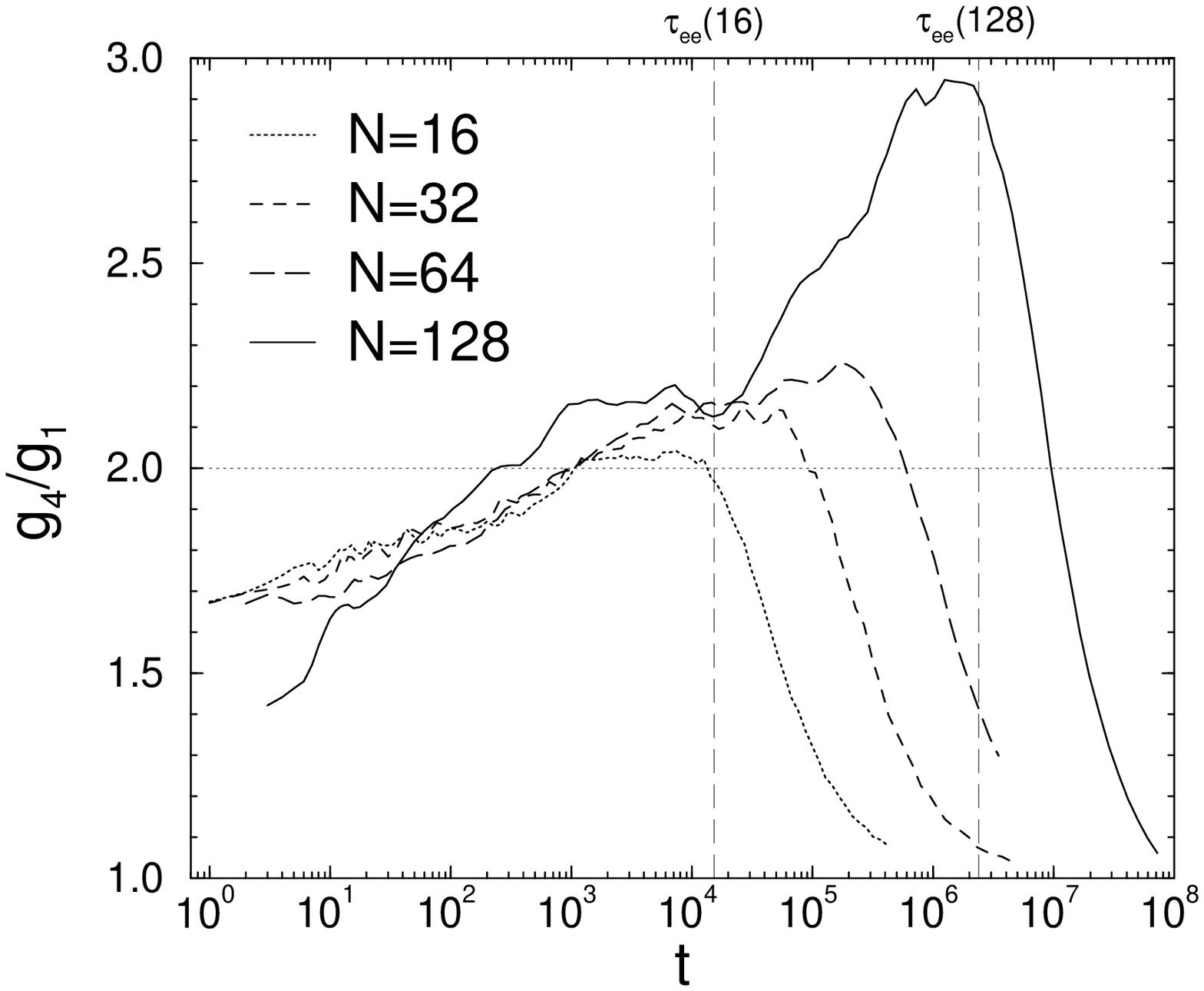}
\caption[]{}
\end{figure}
\begin{figure}
\epsfysize=110mm
\epsffile{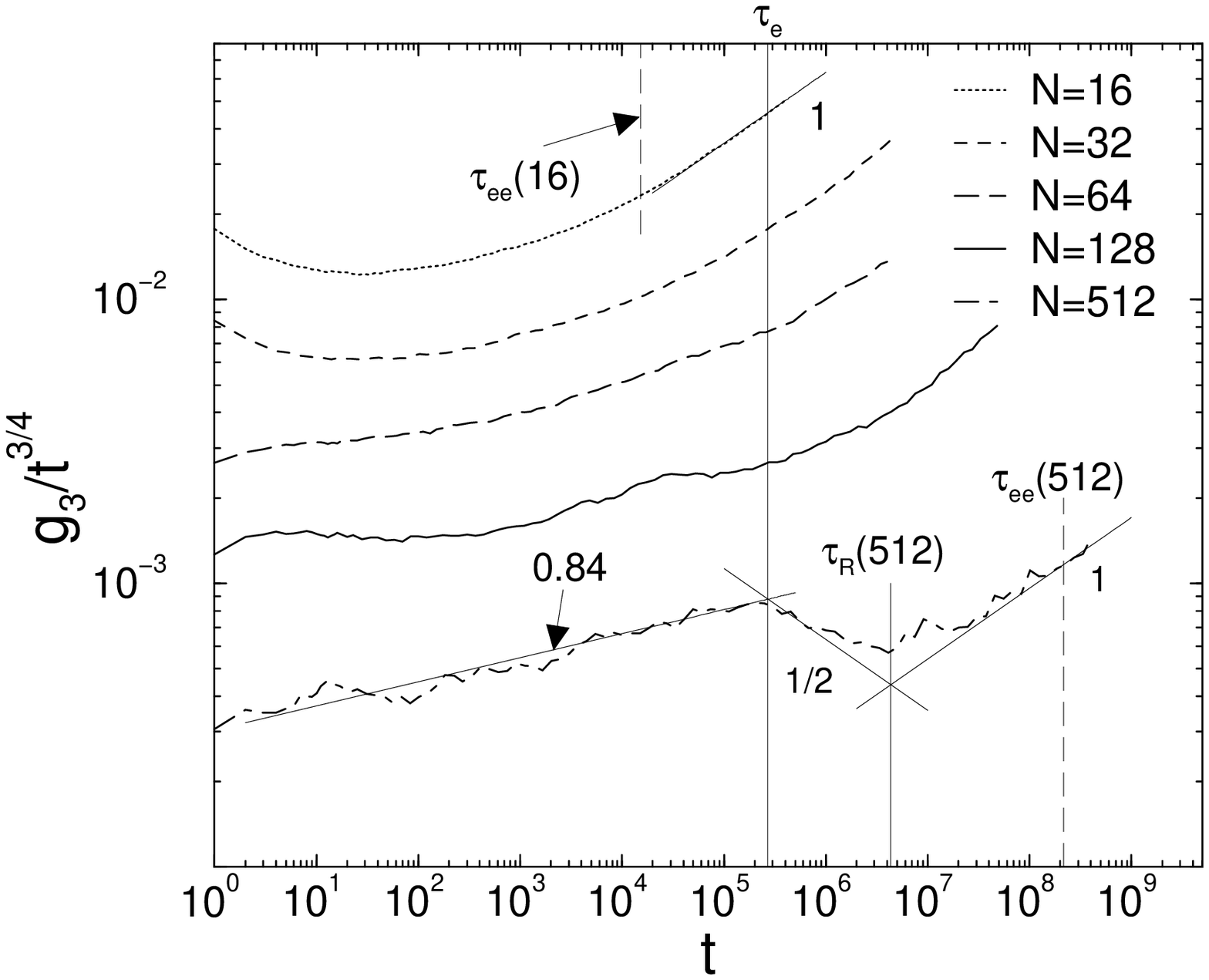}
\caption[]{}
\end{figure}
\begin{figure}
\epsfysize=110mm
\epsffile{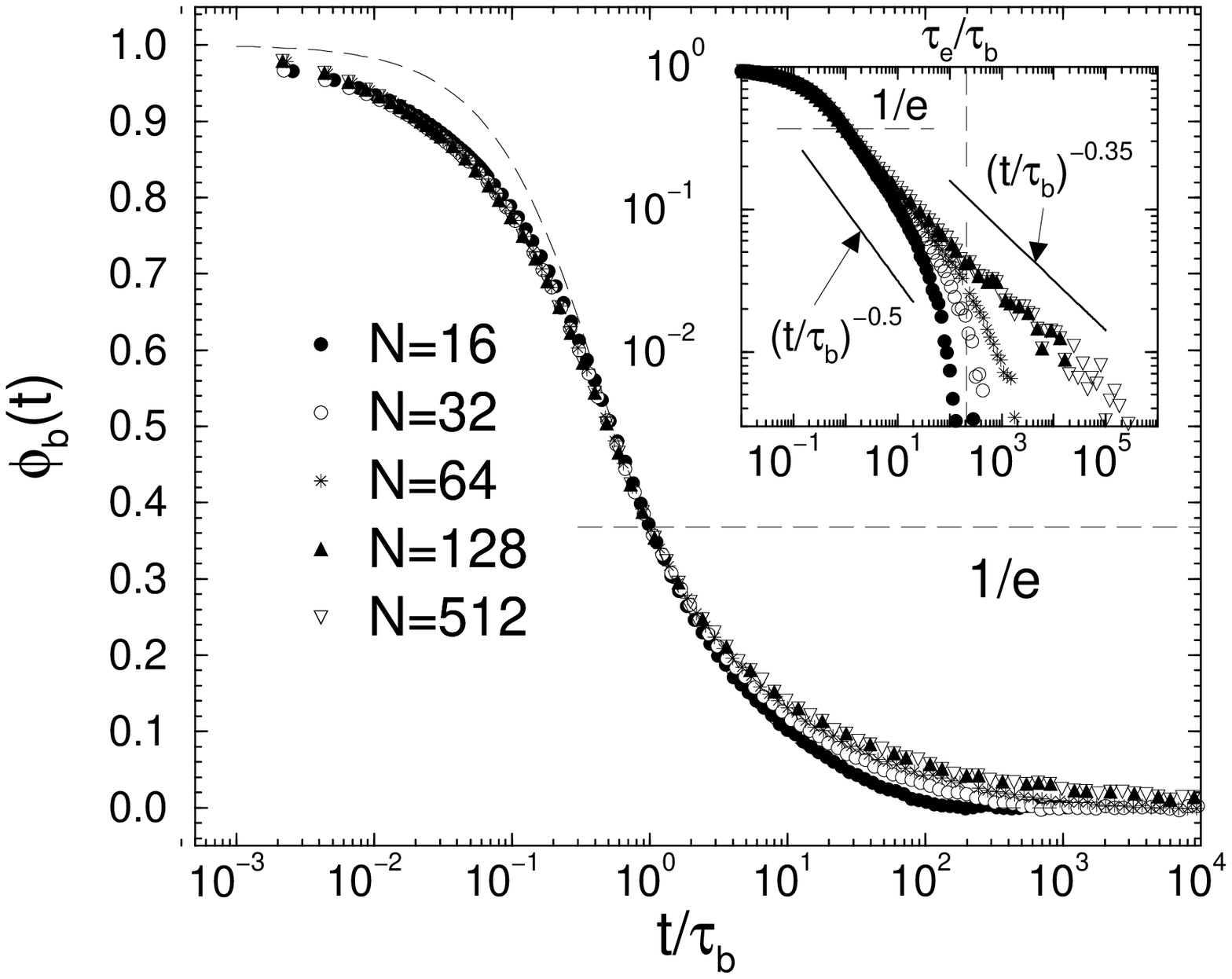}
\caption[]{}
\end{figure}
\begin{figure}
\epsfysize=110mm
\epsffile{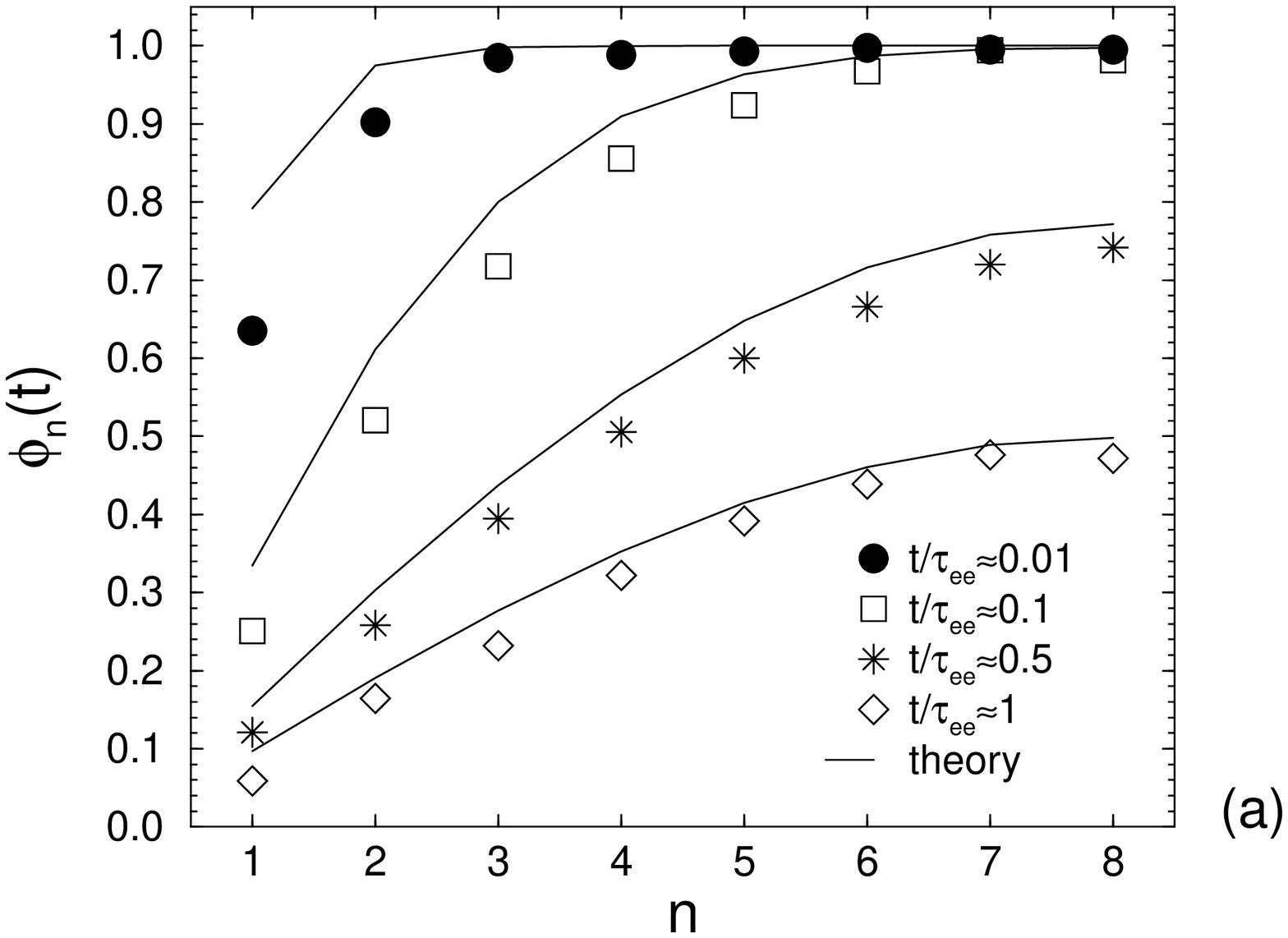}
\epsfysize=110mm
\epsffile{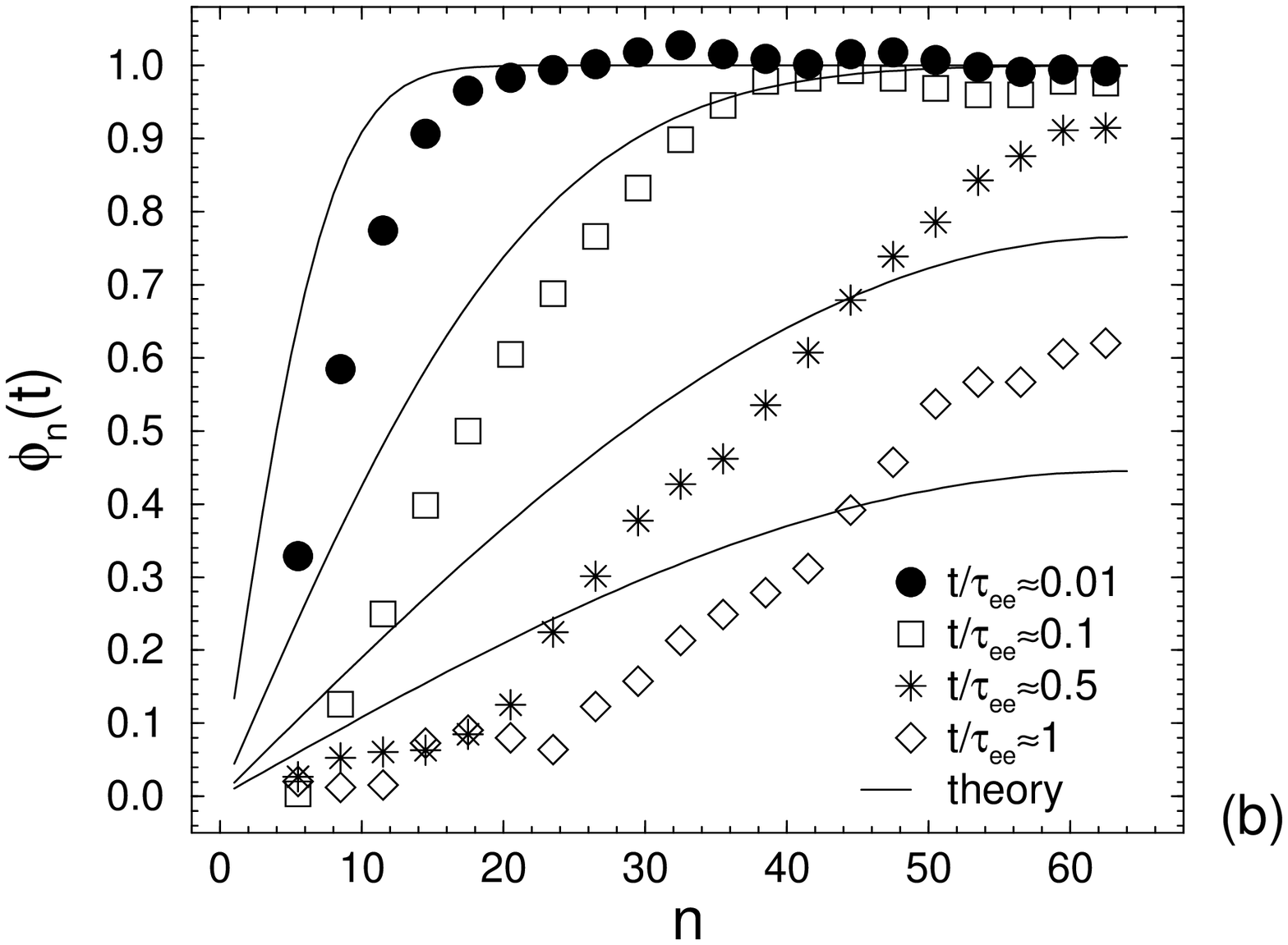}
\caption[]{}
\end{figure}
\begin{figure}
\epsfysize=110mm
\epsffile{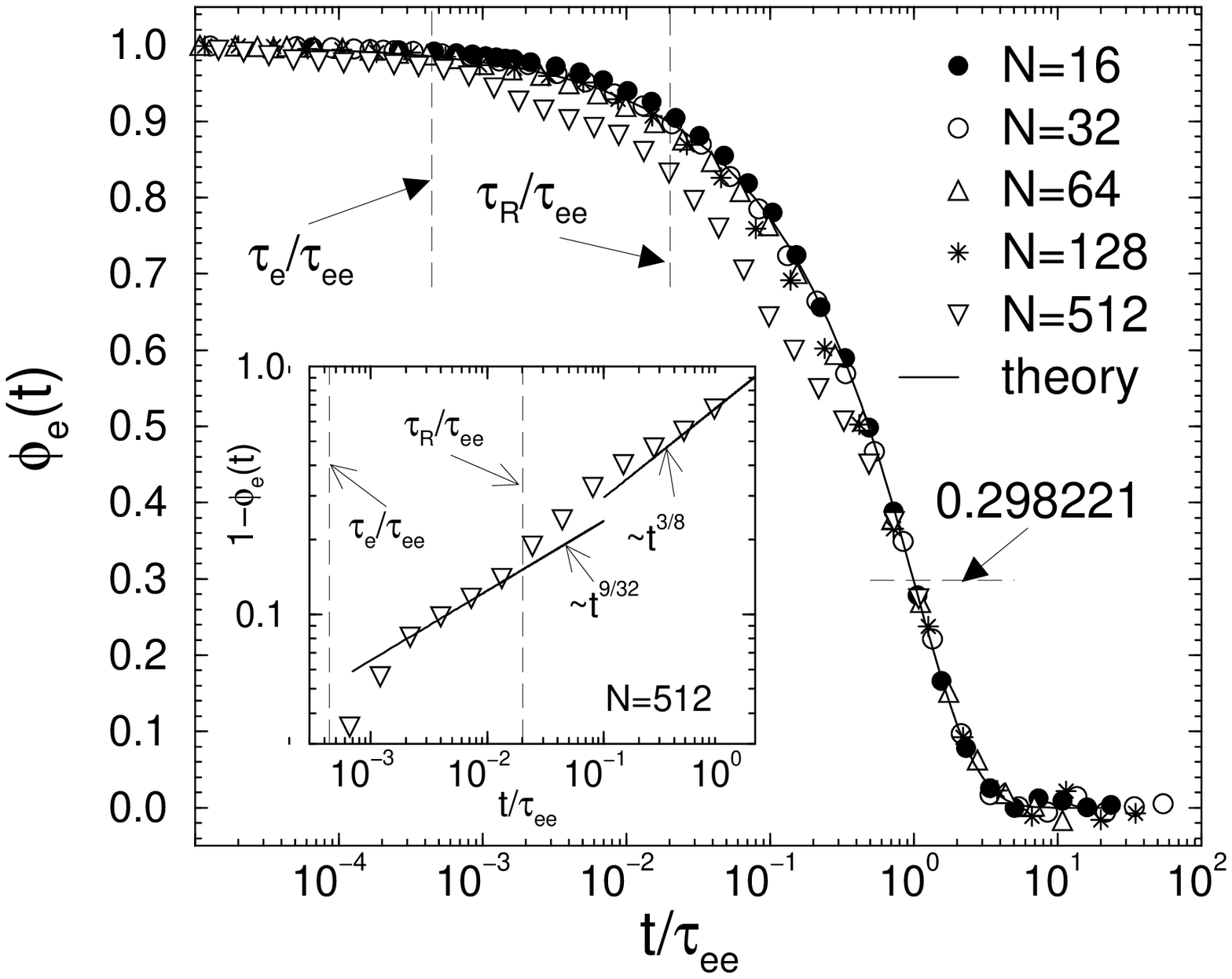}
\caption[]{}
\end{figure}
\begin{figure}
\epsfysize=110mm
\epsffile{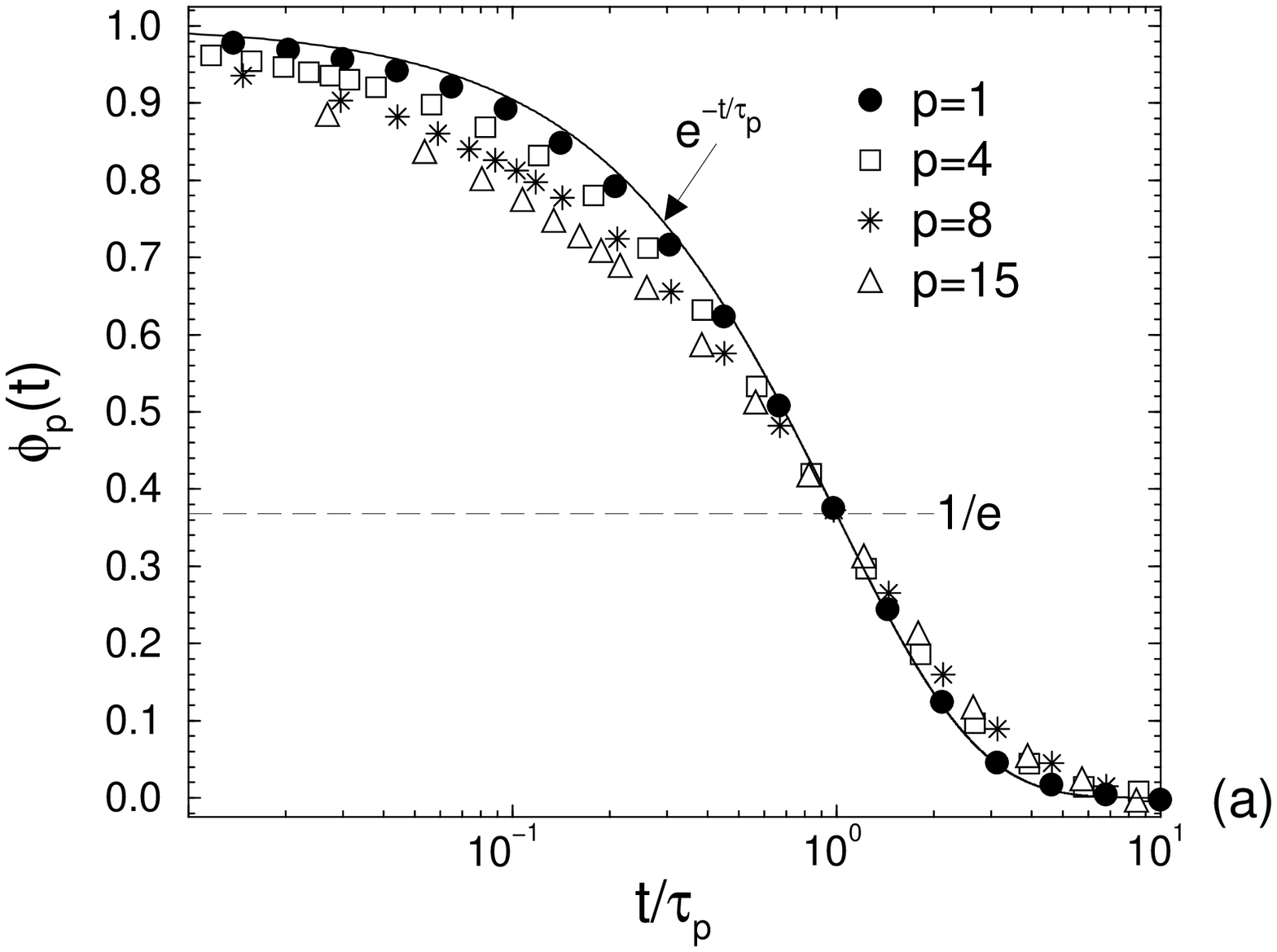}
\epsfysize=110mm
\epsffile{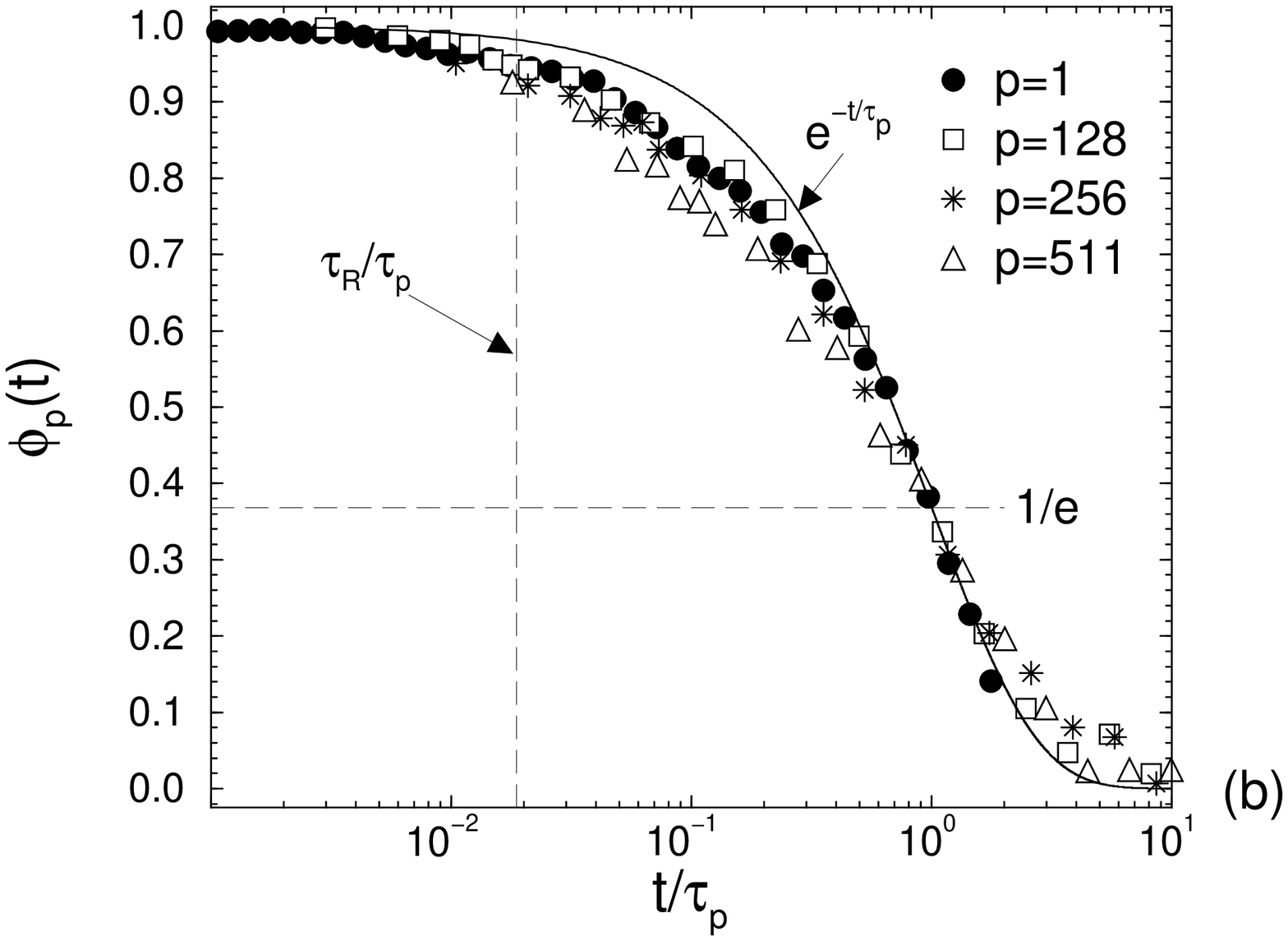}
\caption[]{}
\end{figure}
\begin{figure}
\epsfysize=110mm
\epsffile{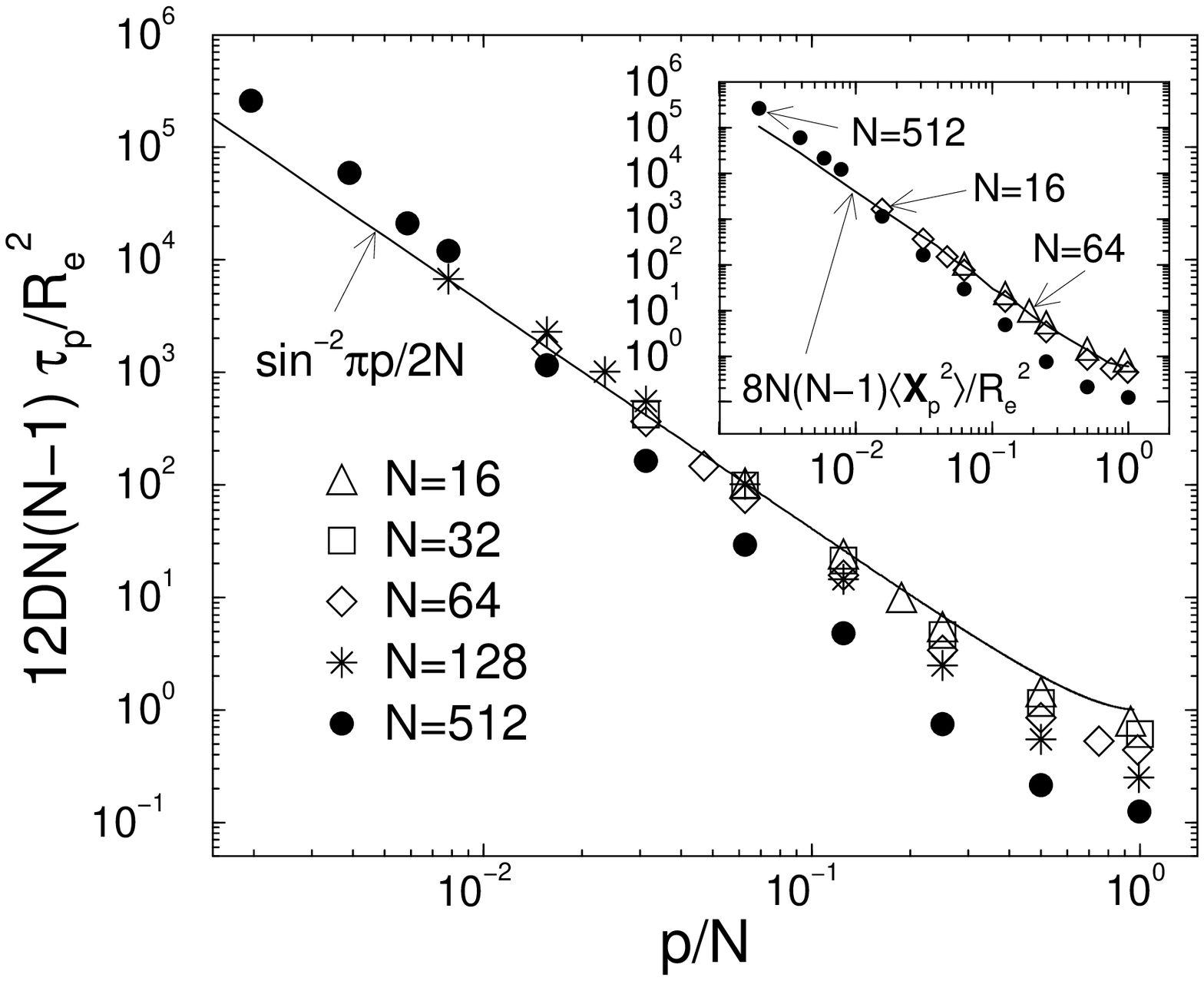}
\caption[]{}
\end{figure}
\begin{figure}
\epsfysize=110mm
\epsffile{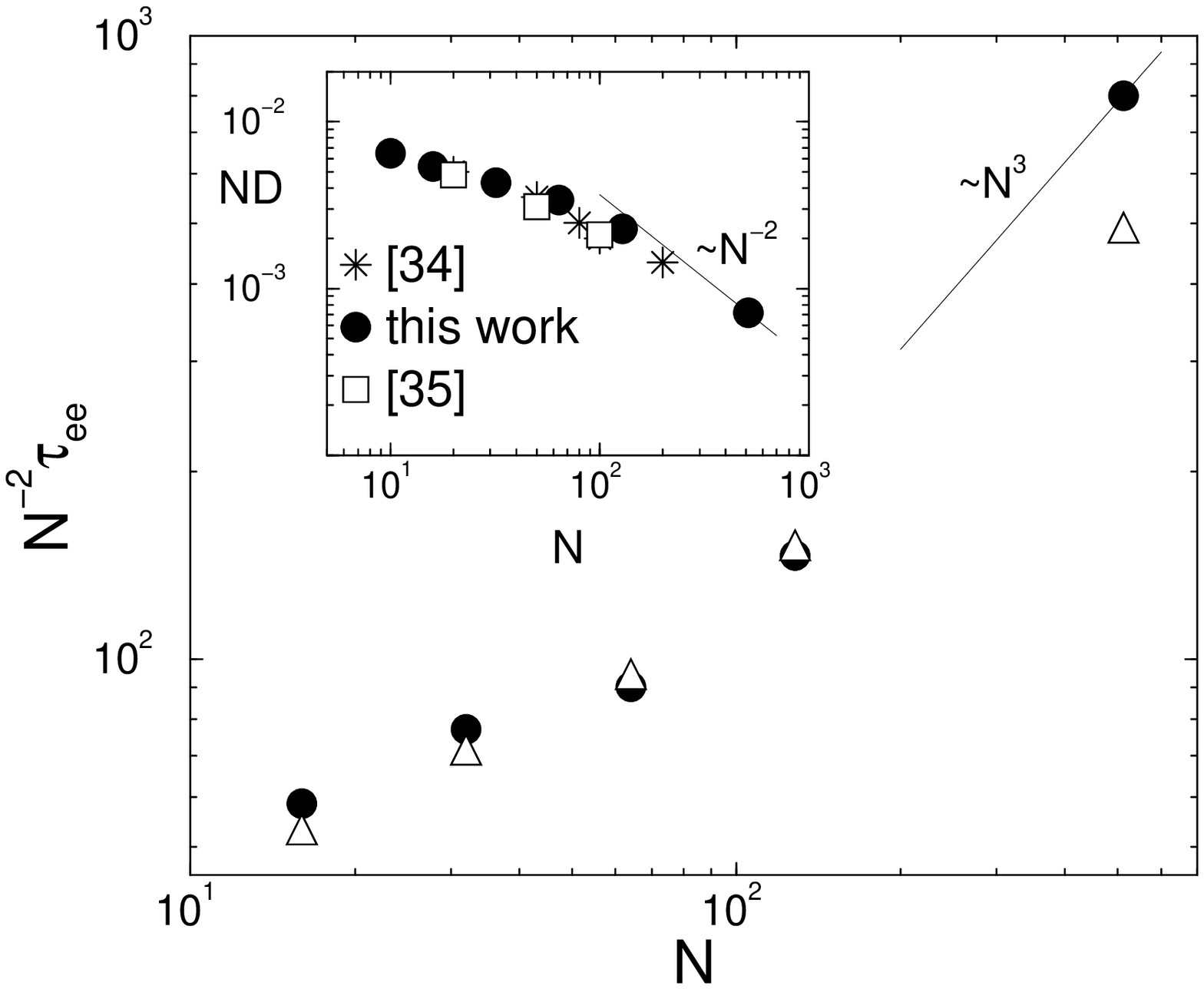}
\caption[]{}
\end{figure}
\end{document}